\documentclass[%
 reprint,
superscriptaddress,
 amsmath,amssymb,
 aps,
pra,
]{revtex4-1}

\usepackage{graphicx}
\usepackage{dcolumn}
\usepackage{bm}
\usepackage{natbib}
\usepackage[export]{adjustbox}
\usepackage{tikz}
\usepackage{appendix}
\usepackage{xcolor}
\usepackage{float}
\graphicspath{Figures}

\usepackage{natbib}
\usepackage[export]{adjustbox}
\usepackage{tikz}
\graphicspath{Figures}

\begin{document}


\title{Shear Melting and Recovery of Crosslinkable Cellulose Nanocrystal-Polymer Gels}

\author{Abhinav Rao}
\affiliation{Department of Mechanical Engineering, Massachusetts Institute of Technology, Cambridge MA 02139, USA
}
 
\author{Thibaut Divoux}
\affiliation{Centre de Recherche Paul Pascal, CNRS UMR 5031 - 115 avenue Schweitzer, 33600 Pessac, France}
\affiliation{MultiScale Material Science for Energy and Environment, UMI 3466
CNRS-MIT,\\ 77 Massachusetts Avenue, Cambridge, Massachusetts 02139, USA}

\author{Gareth H. McKinley}
\affiliation{Department of Mechanical Engineering, Massachusetts Institute of Technology, Cambridge MA 02139, USA
}

\author{A. John Hart}
\email{Corresponding author: ajhart@mit.edu}
\affiliation{Department of Mechanical Engineering, Massachusetts Institute of Technology, Cambridge MA 02139, USA
}


\begin{abstract}
Cellulose nanocrystals (CNC) are naturally-derived nanostructures of growing importance for the production of composites having attractive mechanical properties, and offer improved sustainability over purely petroleum-based alternatives. Fabrication of CNC composites typically involves extrusion of CNC suspensions and gels in a variety of solvents,  in the presence of additives such as polymers and curing agents. However, most studies so far have focused on aqueous CNC gels, yet the behavior of CNC-polymer gels in organic solvents is important to their wider processability. Here, we study the rheological behavior of composite polymer-CNC gels in dimethylformamide, which include additives for both UV and thermal crosslinking. Using rheometry coupled with in-situ infrared spectroscopy, we show that under external shear, CNC-polymer gels display progressive and irreversible failure of the hydrogen bond network that is responsible for their pronounced elastic properties. In the absence of cross-linking additives, the polymer-CNC gels show negligible recovery upon cessation of flow, while the presence of  additives allows the gels to recover via van der Waals interactions. By exploring a broad range of shear history and CNC concentrations, we construct master curves for the temporal evolution of the viscoelastic properties of the polymer-CNC gels, illustrating universality of the observed dynamics with respect to gel composition and flow conditions. We therefore find that polymer-CNC composite gels display a number of the distinctive features of colloidal glasses and, strikingly, that their response to the flow conditions encountered during processing can be tuned by chemical additives.  These findings have implications for processing of dense CNC-polymer composites in solvent casting, 3D printing, and other manufacturing techniques.
\end{abstract}

\maketitle


\section{Introduction}

Colloidal gels are percolated networks of attractive particles and/or polymers \cite{Krall1997, Dickinson1983, Stiakakis2002}. Despite encompassing a wide variety of compositions and architectures, colloidal gels share the common feature of a primarily elastic mechanical response under small strains \cite{Bonnecaze2010, Zaccone2009, Trappe2001}. As such, they are used to design soft solid materials with applications including foodstuff, bio-compatible materials, highly stretchable electronics and soft robotics \cite{Lin2016, Yuk2017, Gibaud2012, Gong2014, Thiele2014, Grindy2015}. Recently, colloidal gels have been employed to produce novel materials via direct ink writing, a versatile additive manufacturing technique in which shear-thinning fluids are deposited by extrusion through a nozzle \cite{Raney2018, Mueller2018, Lewis2004}. Extrusion disrupts the gel's equilibrium microstructure until it yields and reorganizes to accommodate the flow process. The response of the gel after it is subjected to a strong flow depends on the nature of interactions among the colloidal particles. Either the gel accumulates damage under shear until it is irreversibly destroyed (gel collapse), or the gel recovers after flow cessation, and eventually re-forms a percolated network, given a sufficiently long period of rest. These two types of behavior allow to us to distinguish between colloidal gels that experience an irreversible yielding transition 
\cite{Bonn1998, Leocmach2014, Lefranc2014, Keshavarz2017}, and colloidal gels that can be rejuvenated due to the shear-reversible nature of the interparticle attractive interactions such as van der Waals or depletion interactions \cite{Derec2003, Ovarlez2007, Bonn2017, Moghimi2017}. The recovery step of the latter category of gels has been termed ``rheological aging" and the kinetics of such phenomena are a complex function of the volume fraction, the nature of the interparticle interactions and the shear history \cite{Viasnoff2002, Negi2010,Joshi2018, Cipelletti2005, Chaudhuri2017, Bouzid2017}. Being able to tailor the yielding, flow and recovery of colloidal gels is crucial to achieving the desired shape, mechanical integrity and microstructure in direct-write 3D printing and molding.\cite{Yu2001, Lewis2006} Such a task becomes even more challenging when designing colloidal gels as precursors for composite materials, as their rheology strongly depends on the physical and chemical properties of multiple components as well as the chemistry of the solvent. Each of these factors can affect the interparticle interactions and consequently the viscoelastic properties. For instance, the presence of charged species has been shown to alter the liquid to gel transition in laponite suspensions \cite{Smith2016, Angelini2015, Angelini2014}. Suspensions and gels of cellulose nanocrystals are a compelling case in point, as they display a wide variety of solvent-mediated interparticle interactions including van der Waals attraction, hydrogen bonding and electrostatic repulsion \cite{Lagerwall2014, Moon2011}.

Cellulose nanocrystals (CNCs) are rod-like nanostructures consisting of repeating glucose units, and are a subject of growing interest as key ingredients of green composites due to their mechanical and chemical properties, as well as their sustainability and biocompatibility \cite{Habibi2010, Moon2011, Hon1994, Kim2015}. CNCs are used as fillers for a variety of thermoplastics, thermosets and hydrogels, resulting in improved mechanical properties and temperature stability \cite{Fallon2018,Khelifa2016}. Methods used for processing CNC composites include solvent-casting, coagulation and spinning, as well as melt extrusion \cite{Tang2010,Lee2016,BenAzouz2012}. More recently, bulk CNC-based structures have been produced by direct-ink writing \cite{Siqueira2017,Huan2018}. Consequently, the rheological properties of these composites, and especially the interplay of yielding, flow and recovery are strongly connected to our ability to manufacture CNC-based materials with the desired microstructure and properties.

CNCs are most commonly produced by sulfuric acid hydrolysis of wood fibers, which leaves residual sulfate groups on the CNCs resulting in a negative surface charge \cite{Moon2011}. This allows the formation of colloidal CNC suspensions with relatively high concentrations, stabilized by electrostatic repulsion \cite{Abitbol2014}. Moreover, CNCs also form intermolecular hydrogen bonds between surface hydroxyl groups, which results in a sol/gel transition as evidenced by the emergence of a yield stress and a sharp increase in the storage modulus at sufficiently high CNC concentration \cite{Liu2011,Moon2011}. The gel microstructure and concentration at which gelation occurs are dependent on the pH and ionic strength of the solvent, the presence of screening charges, as well as the aspect ratio and polydispersity of the CNCs \cite{10.3389/fmats.2016.00021, Bruckner2016, Way2012, Nordenstrom2017}. The rod-like shape of CNCs and the presence of surface charges also favor the formation of liquid crystal phases, which can give the gels a local crystalline order \cite{Gray2015, Urena-Benavides2011,Lagerwall2014}. The case of aqueous CNC gels has been the most documented: the addition of charge-screening ions results in attractive van der Waals interparticle interactions leading to a broad range of rheological behavior depending on whether the gel is dominated by repulsive or attractive interactions \cite{Xu2018, Dorris2012, Guo2011}. Under external shear, the storage modulus of aqueous CNC gels decreases irreversibly as the gel is subject to increasing strain amplitude, and the CNCs are irreversibly rearranged \cite{Derakhshandeh2013}. A recent study on aqueous CNC gels under extrusion flow visualized by polarized light imaging, has shown that above the yield stress, CNCs display shear-induced alignment that is not trivially controlled by the macroscopic strain applied to the gel \cite{Hausmann2018}. This last result illustrates the complexity of the rheological behavior of CNC gels and calls for a more in-depth investigation into the flows of CNC gels over a wide range of timescales.  Studies on CNC gels have typically been limited to aqueous solvents, whereas the industrial-scale production of CNC/polymer composites requires precursors that are chemically more complex, and which may include oligomers, polymers, and crosslinking agents that necessitate organic solvents. Moreover, the mechanisms driving the microstructural changes associated with yielding in CNC gels have are not thoroughly understood.  

In this article, we study the rheological properties of polymer-CNC gels, in an organic solvent. We first show that these gels experience a progressive and irreversible decrease of their elastic properties under external shear. Using rheology coupled with time-resolved Fourier Transform Infrared Spectroscopy (FTIR), we find that degradation of the elastic properties of the gel is caused by the progressive decrease of the density of hydrogen bonds between the CNCs. Such \textit{shear-melting} of the gel microstructure is irreversible because the hydrogen bonds are not restored after flow cessation. 
By contrast, we show that the addition of relatively small amounts of UV and thermal curing agents to the gel leads to a partially shear-reversible behavior. The UV-curable gels also display a progressive thixotropic decrease of their elastic properties upon external shear, but once the flow is stopped, the gels exhibit a partial recovery of their elastic properties. Applying a series of successive shear-melting steps, and monitoring the viscoelastic recovery of the gels after each step, we explore the dynamics of the gel recovery over a very broad range of cumulative strains. First, we show that the yield stress measured after each period of shear decreases logarithmically with increasing accumulated strain. Second, we reveal how the series of recovery curves, which follow each period of shear, can be scaled onto a universal master curve, providing key insights about evolution in the gel microstructure. While each period of shear leads to an irreversible decrease of the total number of hydrogen bonds, the CNCs also interact via attractive van der Waals interactions (made possible by the curing agent, which plays the role of charge screening) and these attractive interactions are responsible for the partial recovery process. This scenario, which is confirmed by in-situ rheo-FTIR measurements is robust and illustrated over a broad range of CNC concentrations. Finally, we mimic the typical rheological behavior associated with extrusion-based 3D printing of the polymer-CNC gels by performing creep experiments at stresses close to the yield stress. We show that, in the presence of the crosslinking additives, the gels display a rapid recovery following the nonlinear creep flow, suggesting that the curing agent can be used as a control parameter to tune the mechanical integrity of polymer-CNC gels for use in direct ink writing.  

\section{Experimental}

\textit{UV curable CNC gels} (UV-CG) were prepared by first dissolving an epoxide oligomer, a photoinitiator, and a thermal crosslinker in dimethylformamide (DMF). The oligomer is poly(bisphenol-a-co-epichlorohydrin) diglycidyl ether ($M_n\sim355$ g/mol), the cationic photoinitiator is triarylsulfonium hexafluorophosphate (50\% in propylene carbonate) and the thermal crosslinker is 4-aminophenyl sulfone. All reagents were obtained from Sigma Aldrich. Freeze-dried CNC powder (Celluose Lab, New Brunswick, Canada) is added to the mixture amd then dispersed by probe sonication at 40\% amplitude for 15 seconds (750 W Sonics Vibra Cell). Reference \textit{composite gels} (CG) were prepared using the same process, with the exception of adding the photoinitiator and thermal crosslinker. In all the gels, the mass ratio of the epoxide oligomer to the CNCs is 10\%, and the mass of the cationic photoinitiator is 10\% of the mass of the epoxide oligomer. The amine to epoxide molar ratio is 0.4. The mass fractions of the crosslinkers are selected to allow UV curing of millimeter thick layers within a few minutes, and accelerated thermal curing at lower temperatures.

\begin{figure}
    \centering
    \includegraphics[width=0.85\columnwidth]{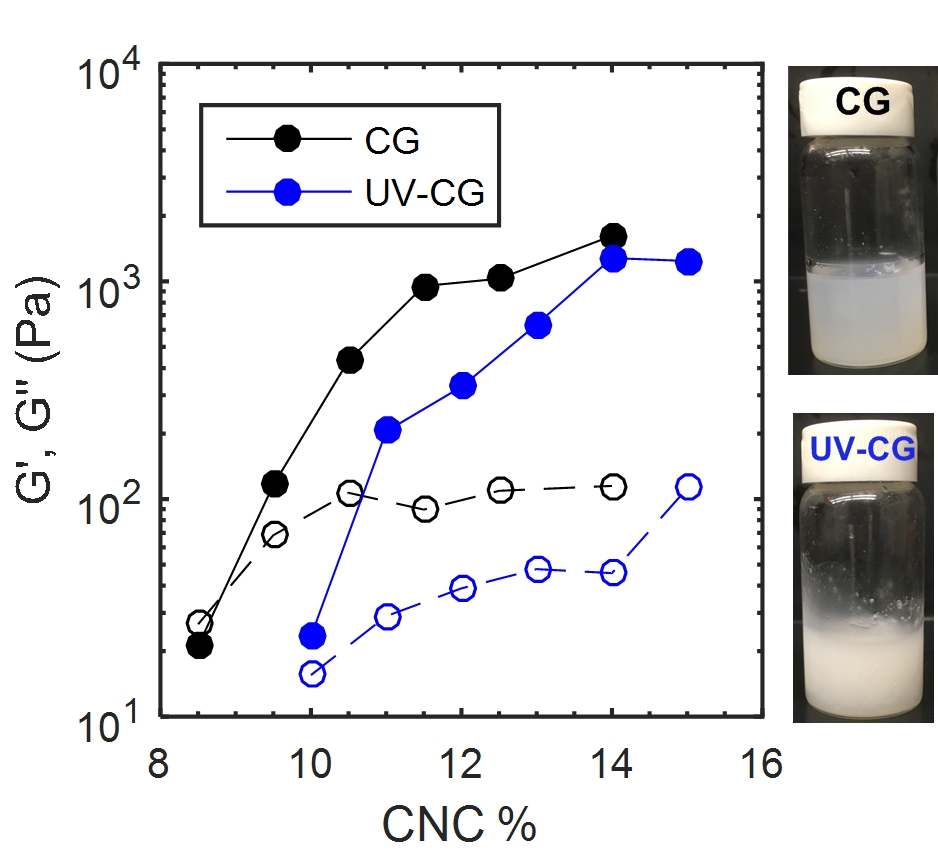}
    \caption{Evolution of the storage and loss modulus, (denoted by $G'$ and $G''$ respectively), for both the composite CNC gel (CG) and UV-curable CNC gel (UV-CG) versus the mass fraction of CNCs. $G'$ is shown with filled symbols, and $G''$ with open symbols. The pristine samples were initially pre-sheared at $\dot \gamma=500$~s$^{-1}$ for 10~s to erase any memory effect due to loading the rheometer. The gels were allowed to recover for 6 minutes after which the values of the viscoelastic moduli were recorded. Insets: photographs of a CNC composite gel (CG) and a UV curable gel (UV-CG).}
    \label{types_of_gels}
\end{figure}

Sonication induces the initial microstructure of the gels, and has been used to break down aggregates of nanoparticles in a variety of colloidal systems, resulting in improved interfacial properties \cite{Mahbubul2014, Nguyen2011}. Here, probe sonication breaks down the freeze-dried CNC powder into smaller nanoparticles, resulting in a large number of hydroxyl (OH) groups available for hydrogen bonding. The average diameter of the CNCs after sonication is around 2.5~nm as measured by dynamic light scattering (see supplementary Fig.~\ref{cnc_size_dls}).  Sonication also generates a complex flow pattern that drives the large-scale association of CNCs efficiently, leading to the formation of the gel. 

Standard rheological tests are performed with a stress-controlled rheometer (AR-G2, TA Instruments) with a 40~mm diameter parallel-plate geometry equipped with a solvent trap. The lower plate (stator),  includes a Peltier module, which is used to maintain the sample temperature at 25$^{\circ}$C. In addition, simultaneous rheology and Fourier-Transform Infrared Spectroscopy (FTIR) measurements are performed with a stress-controlled rheometer (HAAKE MARS, Thermo Fisher) equipped with a Rheonaut module. The rheometer has a 35~mm diameter upper plate (rotor)  and the infrared laser is directed into the sample via an Attenuated Total Reflectance (ATR) crystal of surface area 1~mm$^2$, located in the lower plate of the shear cell (stator). The ATR crystal is positioned at a radial position corresponding to 50\% of the plate diameter, to ensure that the sampled region (which extends between 3 and 10~$\mu$m from the stator) experiences a sufficiently intense local shear flow, while avoiding edge effects. A background IR spectrum is obtained prior to loading each sample, and spectra are collected at 0.1~Hz and synchronized with the rheology data.

\begin{figure}
    \begin{center}
    \includegraphics[width=0.85\columnwidth]{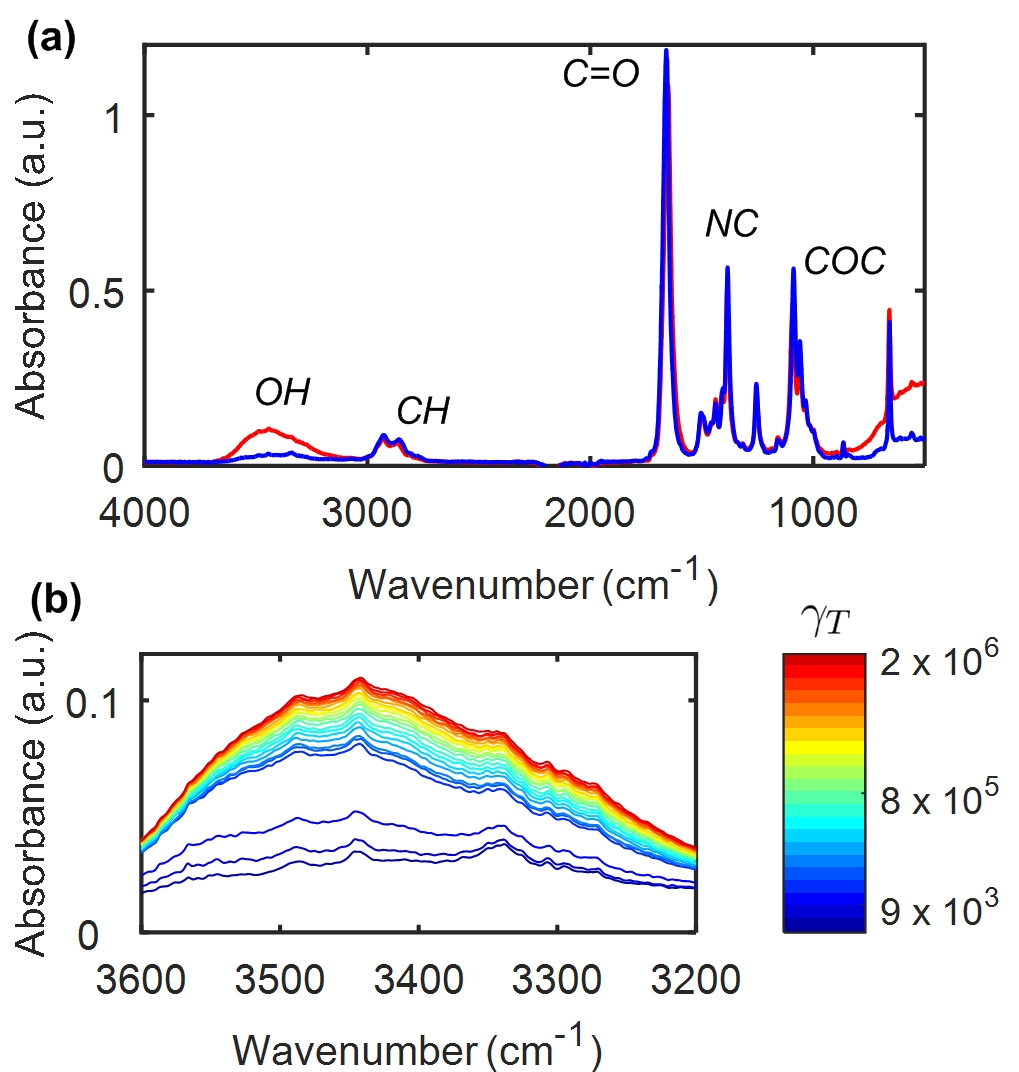}
    \caption{FTIR coupled rheometry of CNC-polymer gels: (a) Mid-IR spectrum of a UV-curable gel with 14\% CNC, acquired on the sample at rest prior to any shear (blue line) and after a large total strain ($\gamma_T \sim 10^6$) accumulated during a creep experiment at $\sigma=105$~Pa (red line). (b) Evolution of the peak corresponding to free hydroxyl groups for different values of the cumulative strain $\gamma_T$}
    \label{UV_CG_IR_spectrum}
    \end{center}
\end{figure}

\begin{figure*}
    \centering
    \includegraphics[width=1.6\columnwidth]{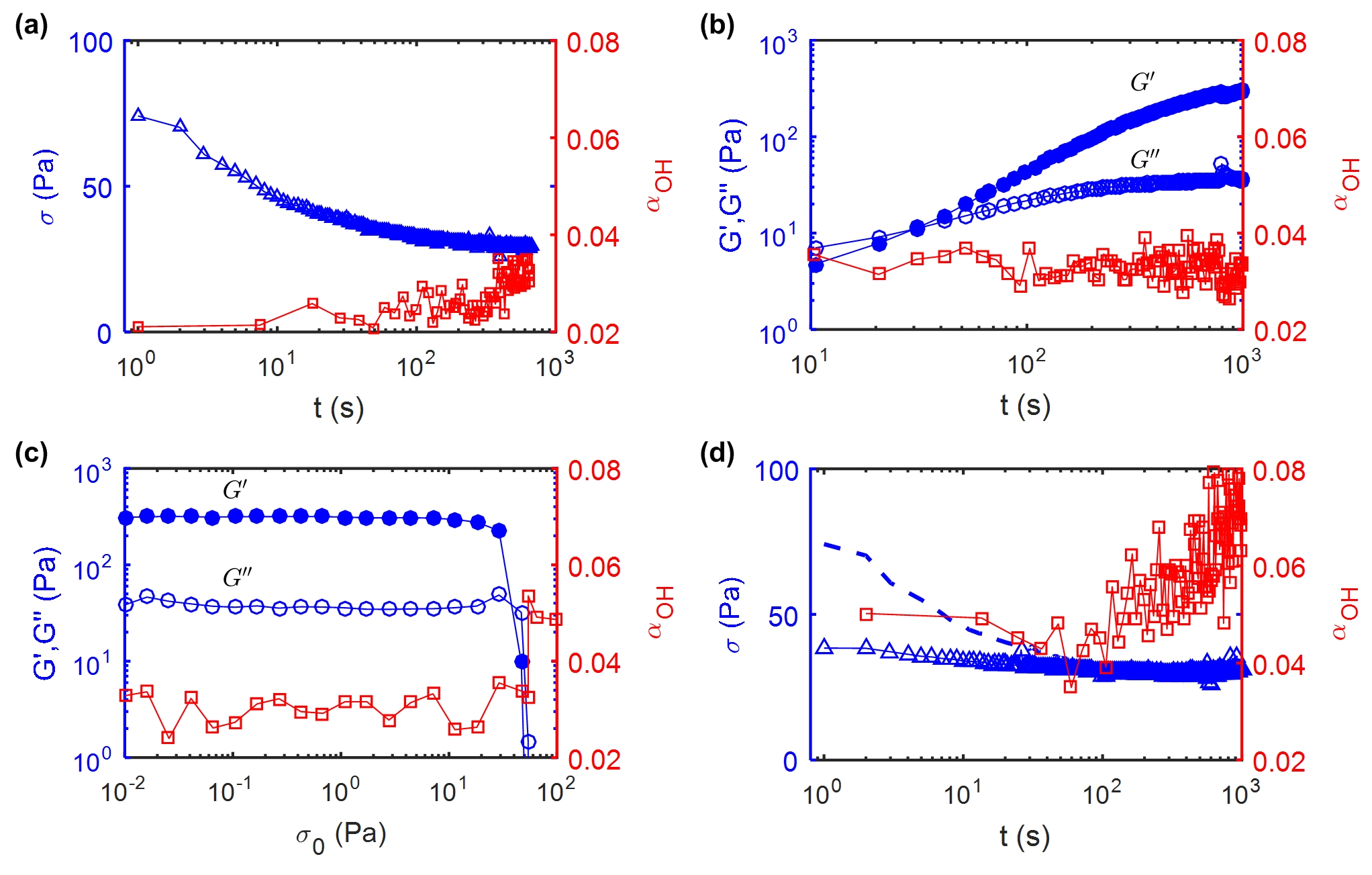}
    \caption{Rheology and simultaneous IR spectroscopy of a 11\% wt.~UV curable gel: (a) Shear-melting step: shear stress $\sigma$ vs time during constant shear flow at $\dot \gamma=500$~s$^{-1}$, (b) Recovery step: evolution of $G'$ (filled symbols) and $G''$ (open symbols) versus time following flow-cessation, (c) Yield stress measurement: oscillatory test following the recovery step, showing the variation of $G'$ and $G''$ versus the stress amplitude, $\sigma_0$ of the oscillations performed at a frequency $f=1$~Hz, and (d) Second shear-melting step: shear stress $\sigma$ versus time. For comparison, the shear stress measured during the first shear-melting step reported in (a) is shown by the dashed line. The right axis of each graph (shown in red) represents the value of $\alpha_{OH}$, which is a measure of the number of free hydroxyl groups inside the gel. Throughout the four successive tests carried out on the same sample, $\alpha_{OH}$ keeps increasing.}
    \label{CSA166_PRT_consolidated}
\end{figure*}

\begin{figure*}[t]
    \begin{center}
    \includegraphics[width=1.6\columnwidth]{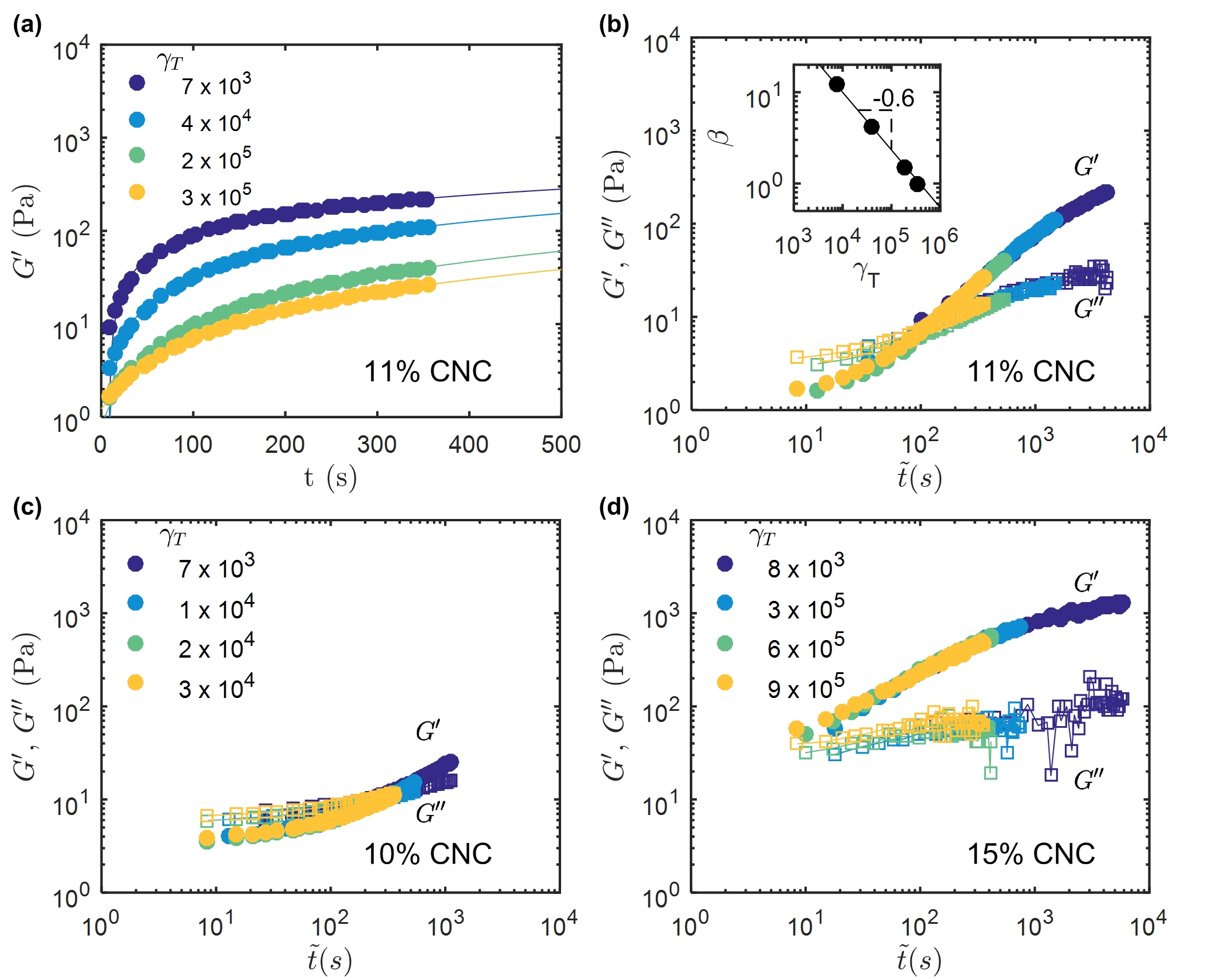}
    \caption{Viscoelastic recovery of UV curable CNC gels: (a) Recovery of the elastic modulus $G'$ versus time after repeated shear-melting steps for a UV curable gel with 11\% wt.~CNC. The cumulative strain, $\gamma_T$ experienced by the sample before each recovery step is indicated on the graph. The fits correspond to the best power-law fits described by Eq.~\ref{power_law_aging}. (b) Master curve for both the elastic modulus $G'$ (filled symbols) and the viscous modulus $G''$ (open symbols) versus a rescaled time axis $\tilde{t}=\beta t$. Inset: Variation of the rescaling parameter $\beta$ vs cumulative strain $\gamma_T$. The line corresponds to a power-law fit $\beta=k\gamma_T^\zeta$ with an exponent $\zeta=-0.63\pm 0.03$ and a prefactor $k=3362$. (c) and (d) show similar master curves to that reported in (b) on the same axes scales, for UV-curable gels with respectively lower and higher CNC content, ie. 10\% wt.~and 15\% wt.~respectively.}
    \label{UV_CG_master_recovery}
    \end{center}
\end{figure*}

\begin{figure*}[t]
    \centering
    \includegraphics[width=1.6\columnwidth]{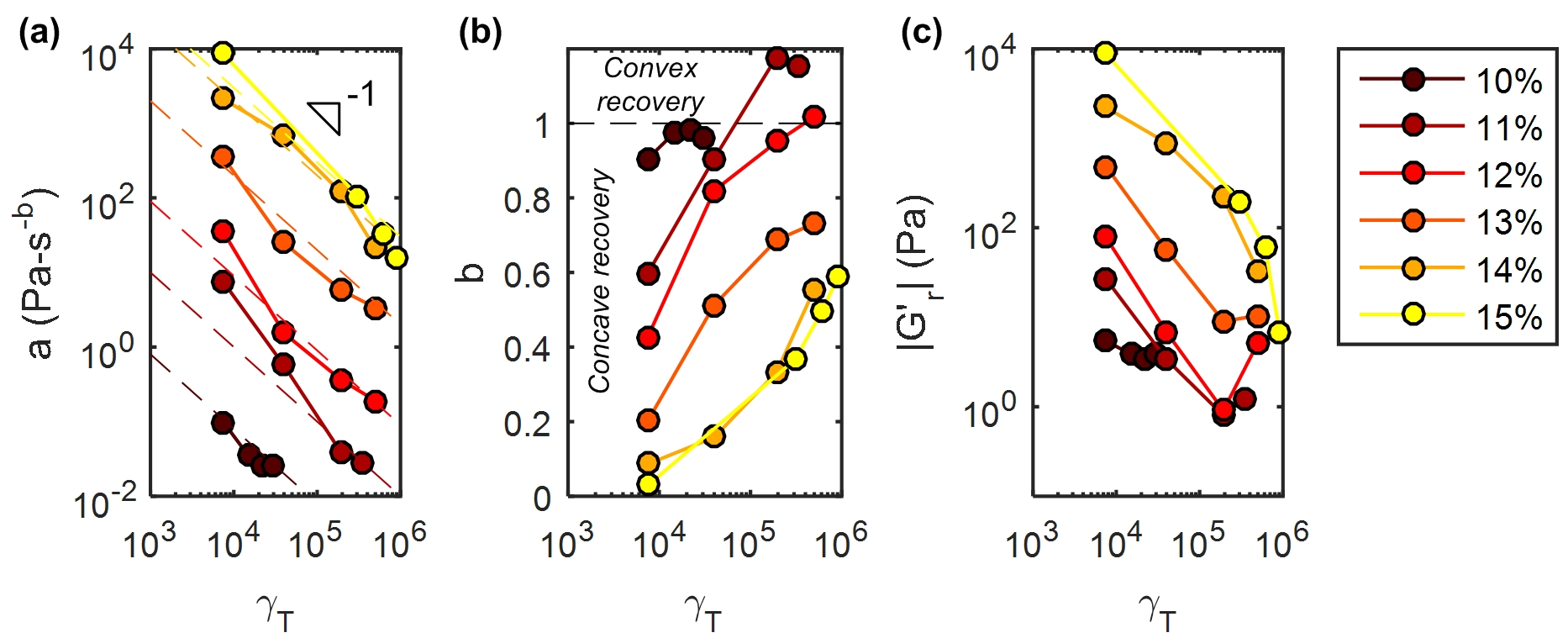}
    \caption{Variation versus cumulative strain of the fit parameters $a$, $b$ and $|G'_r|$, which describe the temporal evolution of $G'$, during the recovery step of UV-curable CNC gels. In (a), the dashed lines are guides for the eye with a slope of -1. In (b), the horizonal dashed line marks the limit between recovery curves showing a concave shape ($b<1$) and those showing a convex shape ($b>1$). Data is shown for UV-curable gels with CNC concentrations ranging between 10\% wt.~and 15\%~wt.}
    \label{UV_CG_recovery_fit_params}
\end{figure*}

The range of CNC concentrations explored was selected such that the polymer-CNC composite always formed a gel when left undisturbed after the sonication step. For CNC content lower than 8\%~wt., we do not observe gelation with the sonication parameters described above, while for CNC content larger than 15\%~wt., the kinetic arrest occurs during the sonication step before the individual CNCs are fully dispersed. Fig.~\ref{types_of_gels} summarizes the changes in the linear viscoelastic properties of the CG and the UV-CG gels, showing the values of the elastic and viscous moduli ($G'$ and $G''$ resp.) as a function of the CNC mass fraction. For both types of composite gels, $G'$ and $G''$ are increasing functions of the CNC mass fraction within the window 8\%~wt.~to 15\%~wt. Moreover, $G'$ saturates at CNC content larger than 11\%~wt. for CG gels (15\%~wt.~for UV-CG gels). This behavior contrasts with the typical power-law--like increase broadly reported for colloidal gels, which involve van der Waals and depletion interactions \cite{Prasad2003, Trappe2001}. For these CNC gels, such saturation may result from the geometric constraints associated with the hydrogen bonding between neighboring CNCs. Indeed, hydrogen bonds behave as localized patchy interactions with limited attraction range.\cite{Nonappa2018} This interpretation is consistent with the finding that the elastic modulus of both types of gels, with and without crosslinkers is limited by the same asymptotic value of $G' \lesssim 2000$~Pa. Comparing the FTIR spectra of pristine UV-curable and composite gels, the effect of the crosslinkers is identified to be physical rather than chemical. The comparison is described in detail in supplementary Fig \ref{uvcg_vs_cg_spectra}. Finally, the physical appearance of the gels are pictured in Fig.~\ref{types_of_gels}. The composite gels are translucent, whereas the UV-curable gels are opaque, indicating that the UV curable gels have a coarser microstructure than the reference composite gels, with a typical size of about a micron. 

FTIR-coupled rheometry allows us to monitor the evolution of the composite gel microstructure under external shear. A representative mid-IR spectrum of a 14\% wt.~UV-curable gel is illustrated in Fig.~\ref{UV_CG_IR_spectrum}a. The most prominent peak between 1600 and 1700~cm$^{-1}$ arises from the carbonyl bond in DMF, while the peak at about $1380$~cm$^{-1}$ is attributed to the NC bond in DMF \cite{Shastri2017}. The peak at about $1100$~cm$^{-1}$ is due to the pyranose ring, which forms the backbone of the CNC. Finally, the broad peak, which spans between 3200 and 3600~cm$^{-1}$ is the signature of the free hydroxyl (OH) groups in the gel \cite{Sofla2016}. This is the primary peak of interest during rheology experiments, as a reduction in the number of hydrogen bonds between the CNCs, results in a growth of the peak intensity due to the increase in concentration of free OH groups \cite{Sofla2016, Kondo1997, Hishikawa2017}.
As an illustrative example, we show in Fig.~\ref{UV_CG_IR_spectrum}b the evolution of this peak for increasing cumulative strain $\gamma_T$ during a creep test at $\sigma=105$~Pa. The dimensionless parameter $\alpha_{OH}$,
\begin{equation}
\alpha_{OH} = \frac{A_{3200-3600}}{A_{1600-1700}}
\label{IR_area_ratio}
\end{equation}
denotes the ratio between the peak area of the hydroxyl groups from the CNCs and that of the carbonyl group from DMF. The baseline of the spectrum between 500 and 4000 cm$^{-1}$ is used in the peak area calculation. The noise in $\alpha_{OH}$ has an amplitude of about 0.006 and the corresponding minimum strain $\Delta \gamma$ required to observe a change in $\alpha_{OH}$ is about 10$^4$.

\begin{figure}
    \centering
    \includegraphics[width=0.85\columnwidth]{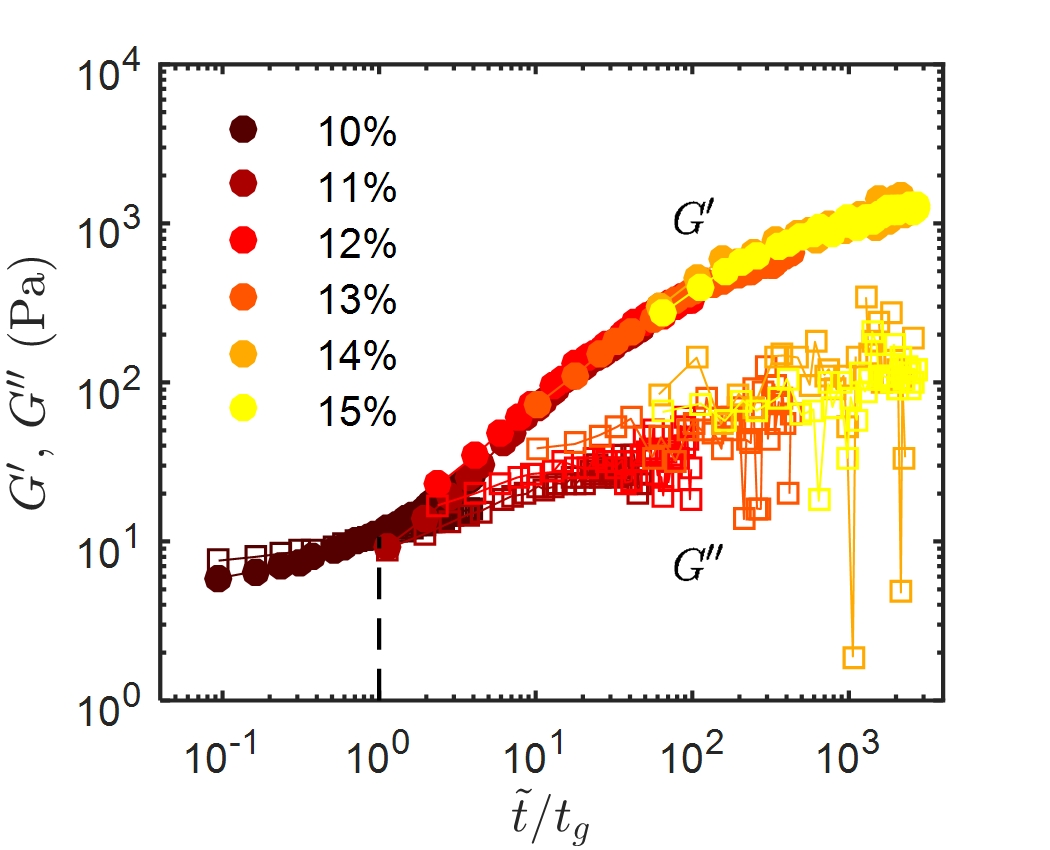}
    \caption{Global master curve of $G'$ and $G''$ versus rescaled time axis $\tilde{t}$, normalized by $t_g$, defined as the crossing point of $G'$ and $G''$. The master curve is obtained from the recovery profiles of $G'$ and $G''$ for UV-curable gels with concentrations ranging from 10\% wt.~to 15\%~wt. The recovery curves for each CNC concentration were recorded at a cumulative strain of $\gamma_T = 7\cdot10^3$. The vertical dashed line indicates the gel point corresponding to $\tilde{t}/t_g=1$.}
    \label{UVCG_concentration_master}
\end{figure}

\section{Results and discussion}

\subsection{Shear melting of UV curable gels}

To quantify the evolution of UV curable CNC gels under external shear, we repeat recursively the same protocol, which consists of three steps: ($i$) a flow at high shear rate to partially degrade the gel microstructure, ($ii$) a period of rest during which we monitor the gel recovery via small amplitude oscillations; and ($iii$) a stress sweep to estimate the yield stress of the gel. We illustrate that sequence in detail in Fig.~\ref{CSA166_PRT_consolidated} for a 11\% wt.~UV curable CNC gel. We first apply a constant shear rate of $\dot{\gamma}=500$~s$^{-1}$ over a fixed duration, here of 660~s but more generally chosen between 10~s and 1000~s so as to control the total strain experienced by the sample. This breaks a fraction of the hydrogen bonds present in the gel. The flow is then stopped by decreasing the shear rate from $\dot{\gamma}=500$~s$^{-1}$ to $\dot{\gamma}=0$~s$^{-1}$ over 10~s. Under constant external shear, the stress $\sigma$, slowly decreases in time, while the structural parameter $\alpha_{OH}$ computed from the IR spectrum displays a progressive increase from 0.02 to 0.035 (Fig.~\ref{CSA166_PRT_consolidated}a).

Following the end of the preshear, the storage and loss modulus of the gel are monitored by small amplitude oscillatory shear for a period of 1000~s (Fig.~\ref{CSA166_PRT_consolidated}b). Right after flow cessation, we find that $G'<G''$, indicating that the gel has been fluidized by the applied shear. After about 30~s, $G'$ becomes larger than $G''$, which indicates that the gel rebuilds, and after 100~s, $G'$ asymptotically increases to a value such that $G'\gg G''$. During this recovery phase, $\alpha_{OH}$ remains constant, $\alpha_{OH}\simeq 0.03$, indicating that the hydrogen bonds broken during the period of high shear do not re-form, and that hydrogen bonding between CNCs cannot be responsible for the elastic recovery of the gel. This is the principal reason why we use the term of ``shear melting" to describe the step of high shear rate. 
To determine the yield stress of the gel after recovery, we perform an oscillatory stress sweep of increasing amplitude from $\sigma_0=10^{-2}$~Pa to 100~Pa (Fig.~\ref{CSA166_PRT_consolidated}c). At low stress values, the gel displays primarily an elastic behavior characterized by $G'\gg G''$, up to the yield point where $G'$ abruptly drops and $G''$ exhibits a local maximum. We use the crossing point of $G'$ and $G''$ to define the yield stress $\sigma_c$ of the gel. Simultaneous measurement of $\alpha_{OH}$ shows that the number of hydrogen bonds remains constant all the way to the yield stress, at which point $\alpha_{OH}$ then increases abruptly to about 0.05, which indicates that the yielding and flow results in a supplemental loss of hydrogen bonds.  

Following the stress-sweep, we repeat for a second time a shear melting step, i.e. a shear flow at $\dot{\gamma}=500$~s$^{-1}$. The evolution of the shear stress and value of $\alpha_{OH}$ versus time during the second shear melting cycle are shown in Fig.~\ref{CSA166_PRT_consolidated}d. For reference, the shear stress measured during the first shear melting step is shown by a dashed line. The initial value of the shear stress in the second cycle is seen to be significantly lower than that measured in the first cycle. Furthermore, $\alpha_{OH}$ increases from 0.05, measured at the end of the stress sweep, to about 0.08. This data is consistent with our hypothesis that the shear-melting and yielding of the UV-curable gel results in an irreversible degradation of some hydrogen bonds in the gel, and that the hydrogen bonds do not re-form even during the extended recovery period.

\begin{figure*}
    \centering
    \includegraphics[width=1.6\columnwidth]{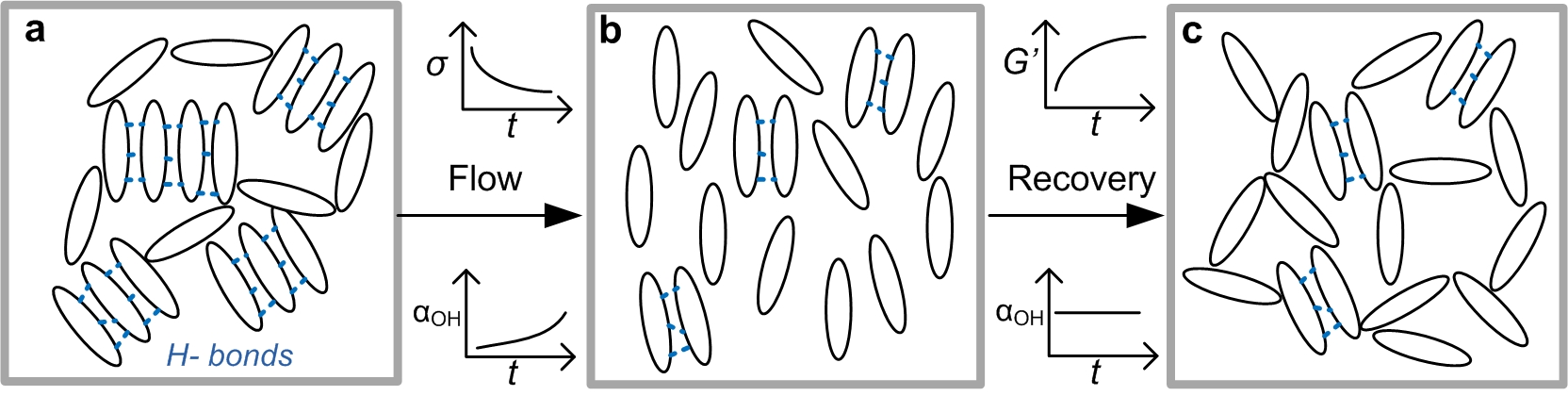}
    \caption{Microstructural evolution during the shear melting and recovery of UV-curable CNC gels: (a) The CNCs in the pristine gel are mostly hydrogen bonded, (b) Shear melting at constant shear rate is accompanied by a monotonic reduction in shear stress, $\sigma$ and an increase in $\alpha_{OH}$. As a result, a fraction of the hydrogen bonds are broken, and the CNCs are free to flow either individually, or in smaller hydrogen-bonded clusters. (c) During the recovery, $G'$ is partially restored, while $\alpha_{OH}$ remains constant, and the CNCs reaggregate by van der Waals forces to form a weaker gel.}
    \label{gel_microstructure}
\end{figure*}

The recovery of the elastic modulus of the UV-curable CNC gels can be attributed to weak van der Waals attractive interactions between the CNCs. Indeed, CNC particles have a residual surface charge but the chemical additives in the UV curable CNC gels inhibit or screen the electrostatic repulsion between the CNCs. The cationic photoinitator is a complex of the form ArS$^+$ PF$_6^-$, where Ar represents a phenyl group. The polyamine curing agent is also polar (Lewis base) and has a structure of the form (R-NH$_2$)$_2$SO$_2$, where R is a hydrocarbon group. The polar curing agents in the UV-curable gels cause an analogous charge screening effect to that of sodium chloride in aqueous CNC suspensions \cite{Lagerwall2014,Xu2018}. Note that the reference composite gels do not show an increase in $G'$ after a sequence of high shear rate, i.e.~they do not recover their elastic properties in the absence of charge-screening (see supplementary Fig.~\ref{CG_PRT_protocol}). This scenario is also consistent with the rapid recovery of the elasticity of the UV-curable gel, since van der Waals attractions are local and lead rapidly to the formation of a percolated network, once the external flow is stopped \cite{Negi2010}.

To further understand the differences in the recovery of these two classes of gels, several cycles of shear melting and recovery were imposed following the protocol described above in Fig.~\ref{CSA166_PRT_consolidated}. 
The temporal evolution of $G'$ during the recovery that follows each shear-melting sequence are reported in Fig.~\ref{UV_CG_master_recovery}a for a 11\% wt.~UV-curable gel (the corresponding evolution in $G''$ is shown in supplementary Fig.~\ref{CSA166_Gpp_recovery}). After each period of high shear, the elastic modulus $G'$ initially increases rapidly, before slowly approaching an asymptotic value. The successive recovery curves however, reach lower $G'$ values for increasing cumulative strain, once again suggesting irreversible degradation during shear-melting steps. The time evolution of $G'$ is well described by the following power-law function: 
\begin{equation}
G'(t) = G'_r + at^b
\label{power_law_aging}
\end{equation}
where $a$ is a measure of the rate of increase of the connectivity of the gel, $b$ is a dimensionless exponent and $G'_r$ is the extrapolation  of $G'$ in the limit of time vanishingly close to the flow cessation. The values of these parameters for gels with different CNC concentrations, are reported in Fig.~\ref{UV_CG_recovery_fit_params}, and display monotonic trends with increasing cumulative strain, $\gamma_T$. First, the rate parameter, $a$ decreases as a power-law with increasing $\gamma_T$ (Fig.~\ref{UV_CG_recovery_fit_params}a). Second, $b$, which captures the curvature of the recovery curve, increases as a logarithm of $\gamma_T$ (Fig.~\ref{UV_CG_recovery_fit_params}b).  
Finally, $|G'_r|$ decreases as a power-law with increasing $\gamma_T$ (Fig.~\ref{UV_CG_recovery_fit_params}c). The CNC concentration impacts the absolute value of these parameters, but only weakly affects the logarithmic dependence with $\gamma_T$. Such a robustness with composition suggests that the gel recovery displays a generic behavior that should be captured by a simple rescaling. 

Indeed, all the recovery curves presented in Fig.~\ref{UV_CG_master_recovery}a, which are determined after various cumulative strains, $\gamma_T$, can be rescaled into a single master curve in a logarithmic representation, by translating the recovery curves along the horizontal (time) axis by introducing the following timescale $\tilde{t} = \beta t$, where $\beta$ is a scaling factor. The same rescaling factor is also used for $G''$. Fig.~\ref{UV_CG_master_recovery}b shows the master recovery curve for the 11\% wt.~CNC gel, and the corresponding rescaling factor $\beta$ decreases as a power-law with increasing $\gamma_{T}$ (inset in Fig.~\ref{UV_CG_master_recovery}b). 
At low $\gamma_T$, the gel recovers rapidly and $G'$ exceeds $G''$ immediately following the shear-melting step. As $\gamma_T$ increases, the recovery dynamics of $G'$ and $G''$ trace the master curve, and the rate of recovery becomes progressively slower. At large cumulative strain, $G'$ is observed to be smaller than $G''$ immediately after flow-cessation, and to recover slowly thereafter. This observation is robust and master recovery curves were also determined for UV-curable gels with 10\% wt.~and 15\%~wt.~CNCs (Fig.~\ref{UV_CG_master_recovery}c and \ref{UV_CG_master_recovery}d respectively). The rescaling factors for the 10\%~wt. and  15\%~wt. gels also follow similar power laws to the 11\%~wt. gel, and are shown in supplementary Fig.~\ref{Beta_vs_gammaT}. At low CNC concentration (10\% wt.), a lower strain is required to fluidize the gel and the recovery is significantly slower. Consistently, at higher CNC concentration (15\% wt.), the elasticity of the gel recovers much faster, with $G'>G''$ almost instantly upon flow cessation for the entire range of cumulative strain explored. Furthermore, the recovery profiles obtained for UV-curable gels with different concentrations, after the same shear history ($\gamma_T = 7 \cdot 10^3$), can all be combined into a global master curve by normalizing the time axis by a characteristic time, $t_g$, defined as the crossing point of $G'$ and $G''$ (Fig.~\ref{UVCG_concentration_master}). This sets a common reference point for all the recovery curves, regardless of the CNC concentration of the gel. The global master curve illustrates the universality of the shear-melting and recovery behavior of the UV-curable gels under non-linear external shear. 

Combining the insights from in-situ FTIR measurements and rheology, we can now discuss the mechanism of shear-melting and recovery, as well as illustrate the universality of the gel recovery in this shear-sensitive, weakly attractive system, with respect to cumulative strain and CNC concentration. Fig.~\ref{gel_microstructure} shows a schematic of the postulated microstructural evolution during shear melting and recovery. The gel is initially composed mostly of hydrogen-bonded clusters of CNCs (Fig.~\ref{gel_microstructure}a). As we increase the strain applied to the composite gel, an increasing number of hydrogen bonds are irreversibly broken, and more CNCs are free to be rearranged by the local flow field (Fig.~\ref{gel_microstructure}b). Following each shear-melting step, the gel partially recovers at a rate that decreases with increasing accumulated strain. This result suggests that the strongly hydrogen-bonded network of CNCs is progressively replaced by clusters of CNCs that interact via weak van der Waals attractive forces (Fig.~\ref{gel_microstructure}c). A striking feature seen in the recovery of the 11\% wt.~gel is that $G'$ is initially convex at low $\gamma_T$ before transitioning to a concave shape at large $\gamma_T$ (Fig.~\ref{UV_CG_master_recovery}b). The recovery exponent, $b$, is less than 1 at small strains ($\gamma_T = 7\cdot10^3$) and greater than 1 at larger strains (Fig.~\ref{UV_CG_recovery_fit_params}b). At small strains, the concave recovery curve reflects the rapid initial recovery of the CNCs, which are trapped in local ``cages`` formed by their neighbors. As the accumulated strain increases, the shape of the viscoelastic recovery becomes increasingly convex, indicating that the recovery is initially slow, but accelerates after a quiescent period. Correspondingly, $G''$ is initially lower than $G'$ suggesting that the gel is fluidized, permitting the CNCs to flow either individually or in clusters. Combining these observations, we deduce that recovery after the application of large strains may first involve hydrodynamic rearrangements of the CNCs, followed by rapid percolation of van der Waals bonded clusters of CNCs.

These distinctions in the dynamics of viscoelastic recovery are also observed by comparing gels of different CNC concentrations after similar shear histories. On the one hand, the 10\% wt.~CNC gel breaks down at low strains, with a recovery exponent $b \sim 1$, and a recovery rate $a \ll 1$, indicative of very slow recovery. On the other hand, the 15\% wt.~CNC gel exhibits a highly concave ($b<1$) recovery profile at low strains. While $b$ approaches 1 at higher strains, $a$ is about 4 orders of magnitude larger than that observed for the 10\% wt. gel, suggesting a more rapid recovery. Moreover, the initial portion of the recovery of the 15\% wt. gel remains rapid even at high strains, most likely because the loading of CNCs is sufficient for rapid clustering to take place via van der Waals interactions. As a result, the initial percolation takes place too rapidly in the 15\% wt.~gel to be measured experimentally, and $G' > G''$ at all times greater than 8.5s (the first recorded data point) during the recovery.

Further emphasizing the universality of the recovery dynamics, the rate parameter $a$ can be directly compared across gels with different CNC concentrations without normalization, although the final value of $G'$ after recovery increases with CNC concentration. This is because the value of $G'$ corresponding to the crossover of $G'$ and $G''$ during recovery is almost identical for all the gels, about 20~Pa. This indicates that all the gels pass through a similar minimal percolated state during the recovery step. While the cumulative shearing strain $\gamma_T$ increases from $7\cdot10^3$ to $5\cdot10^5$, $a$ decreases monotonically from 7.5 to 0.02 for the 11\% wt.~gel. With increasing $\gamma_T$, more hydrogen bonds are broken allowing hydrodynamic re-ordering of the CNCs, and resulting in smaller clusters. The dependence of the aggregation rate on cluster size is widely observed in attractive colloidal gels.\cite{Cipelletti2000} Illustrating that lowering the CNC concentration has a similar effect to increasing $\gamma_T$, the initial rate of recovery at $\gamma_T = 8\cdot10^3$, is 0.1~Pa~s$^{-b}$ for a 10\% wt.~CNC gel and 8500~Pa~s$^{-b}$ for a 15\% wt.~CNC gel. 
An analogous regulation of recovery behavior is seen in aging silica gels, where temperature, rather than the accumulated strain, alters the dynamics of particle aggregation following shear melting \cite{Negi2014}. More generally, the universality we demonstrate via the master curves in Figs.~\ref{UV_CG_master_recovery} and \ref{UVCG_concentration_master} is consistent with the theory that fluidization of a colloidal glass by shear is analogous to decreasing the colloidal volume fraction or increasing the thermal energy of the system\cite{Eisenmann2010}.

\begin{figure}
    \centering
    \includegraphics[width=0.9\columnwidth]{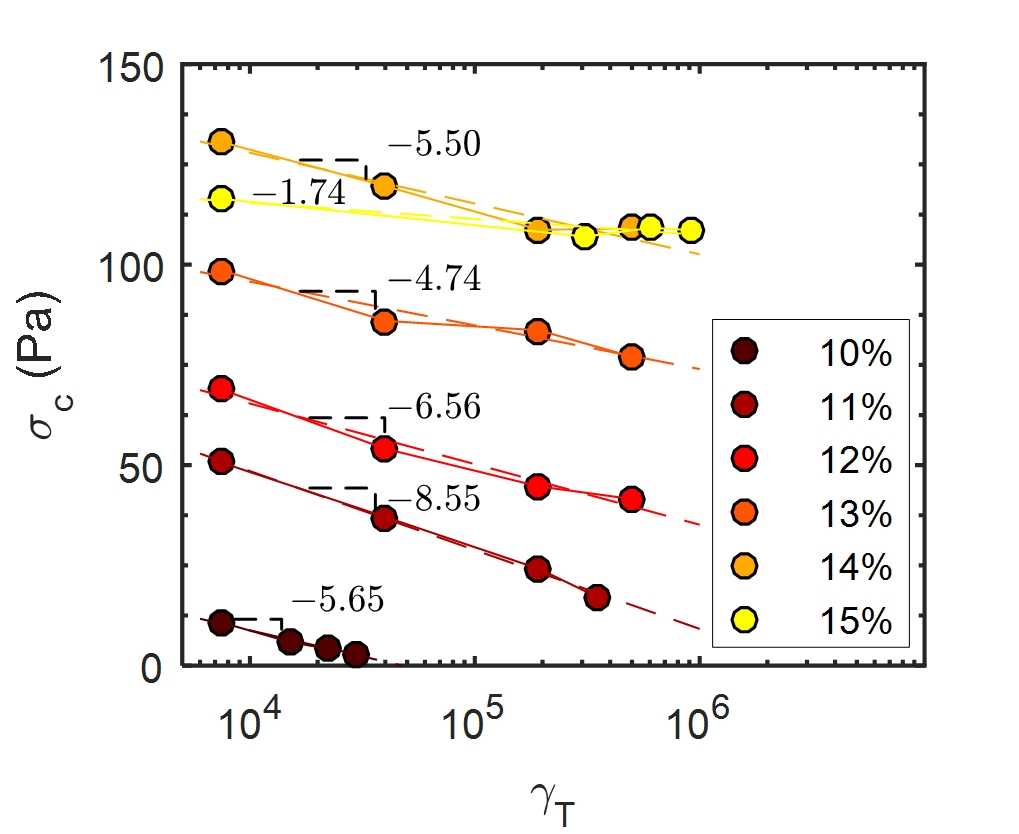}
    \caption{Yield stress, $\sigma_c$, of UV-curable CNC gels for different CNC concentrations ranging between 10\% and 15\% wt.~versus cumulative strain $\gamma_T$. Each yield stress  was obtained by performing a stress sweep at a frequency of $f=1$~Hz. See Fig.~\ref{CSA166_PRT_consolidated}(c) for the measurement of the first point of the 11\% wt.~data set. The dashed lines correspond to the best fit of the data with a logarithmic function: $\sigma_c=\sigma_c^{(0)}+B\ln(\gamma_T)$, where $\sigma_c^{(0)}$ and $B$ serve as free parameters. The fit parameter $B$, which is reported on the graph, does not show any trend with the concentration of CNCs.}
    \label{UV_CG_yield_stress}
\end{figure}

The evolution of the yield stress, $\sigma_c$ (Fig.~\ref{UV_CG_yield_stress}) with cumulative strain for the UV-curable gels, was determined by performing a stress sweep of the form shown in Fig.~\ref{CSA166_PRT_consolidated}c following each recovery step. The yield stress is defined as the minimum applied stress amplitude for which $G'<G''$. The initial yield stress of the gel increases with increasing CNC concentration, except for 14\% wt.~and 15\% wt.~gels, which display a similar yield stress. This result is in agreement with the existence of a plateau in the dependence of the elastic modulus with the CNC concentration, for weight fractions larger than 14\% (Fig.~\ref{types_of_gels}).  
For increasing $\gamma_T$, the UV-curable gels display a monotonic decrease in the yield stress following a logarithmic dependence in $\gamma_T$. For the lowest concentration explored (10\%~wt.), the gel becomes fully fluidized, i.e.~it no longer exhibits a yield stress for a cumulative strain of $\gamma_T\simeq 3\cdot10^4$. 
Note that shear-induced irreversible rearrangements of CNCs were reported in aqueous suspensions using light scattering techniques \cite{Derakhshandeh2013}. However, we demonstrate for the first time that the irreversibility of the yielding process is linked to the loss of hydrogen bonds between the CNCs, and that partial recovery of elastic properties of the gels is made possible by the charge-screening effect of the crosslinkers, leading to the formation of a percolated network mediated by van der Waals attractions.

\begin{figure*}
    \begin{center}
    \includegraphics[width=1.7\columnwidth]{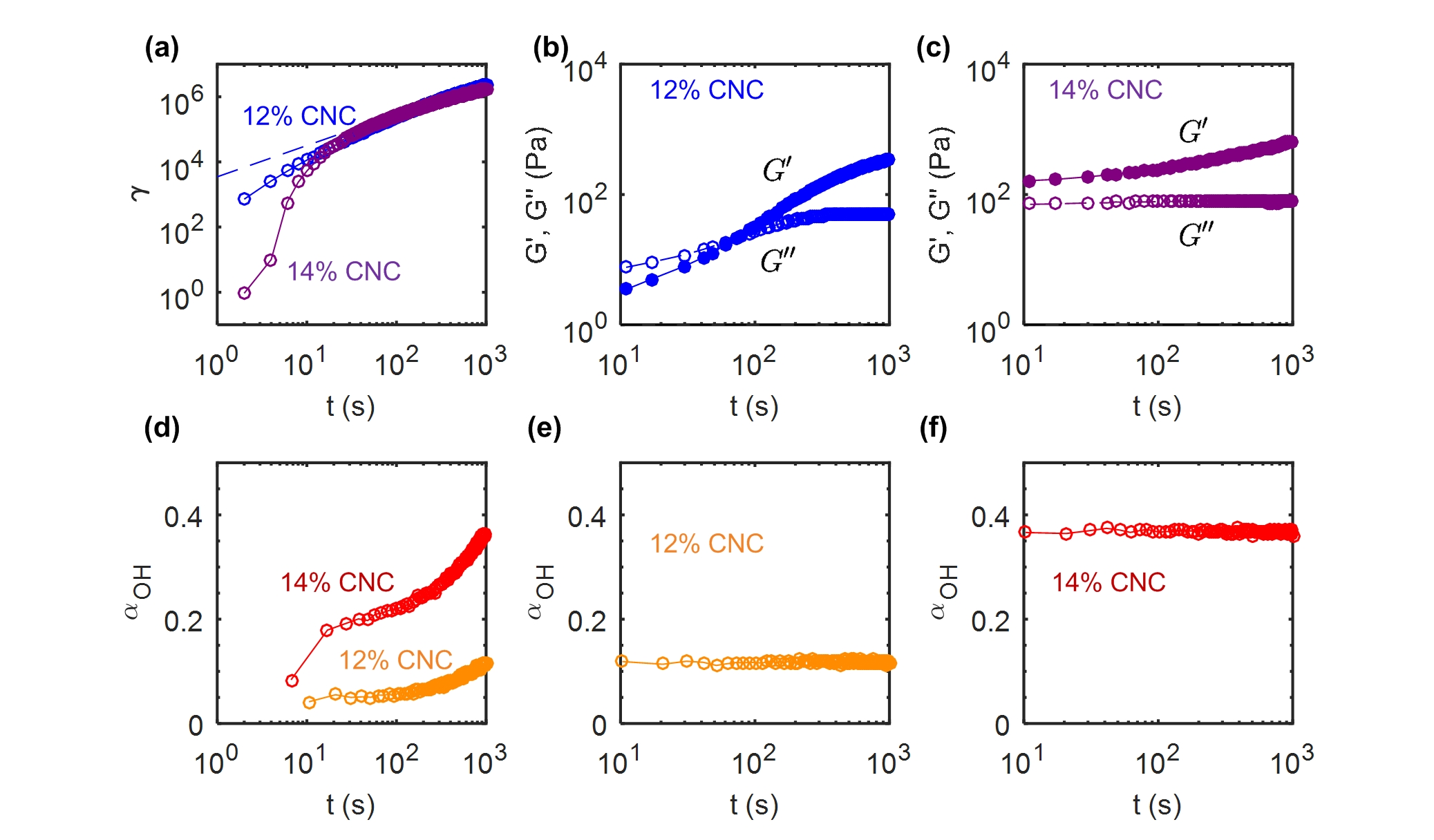}
    \caption{Creep and recovery of UV curable (UV-CG) gels: (a) Shear strain $\gamma$ versus time during creep flow at $\sigma=65$~Pa for the 12\%~wt.~gel and $\sigma=105$~Pa for the 14\%~wt. CNC gel. These stress values were chosen to be close to the respective yield stresses, ($\sigma/\sigma_c = 0.99$) of each gel (see supplementary Fig.~\ref{UVCG_CG_yield_creep} for the determination of the yield stress). The dashed line corresponds to the best power-law fit of the data $\gamma(t) \propto t^{0.96}$ for the 12\% wt.~gel. (b) and (c) Elastic and viscous modulus versus time, measured by small amplitude oscillatory shear of amplitude $\sigma_0=0.5$~Pa and frequency $f=1$~Hz for the 12\%~wt. and the 14\%~wt.~gel respectively. Evolution of the IR peak ratio parameter $\alpha_{OH}$ versus time during (d) the creep flow and (e)--(f) the recovery of the 12\%~wt.~and 14\%~wt.~gels.}
    \label{UV_CG_creep}
    \end{center}
\end{figure*}

\begin{figure*}
    \begin{center}
    \includegraphics[width=1.7\columnwidth]{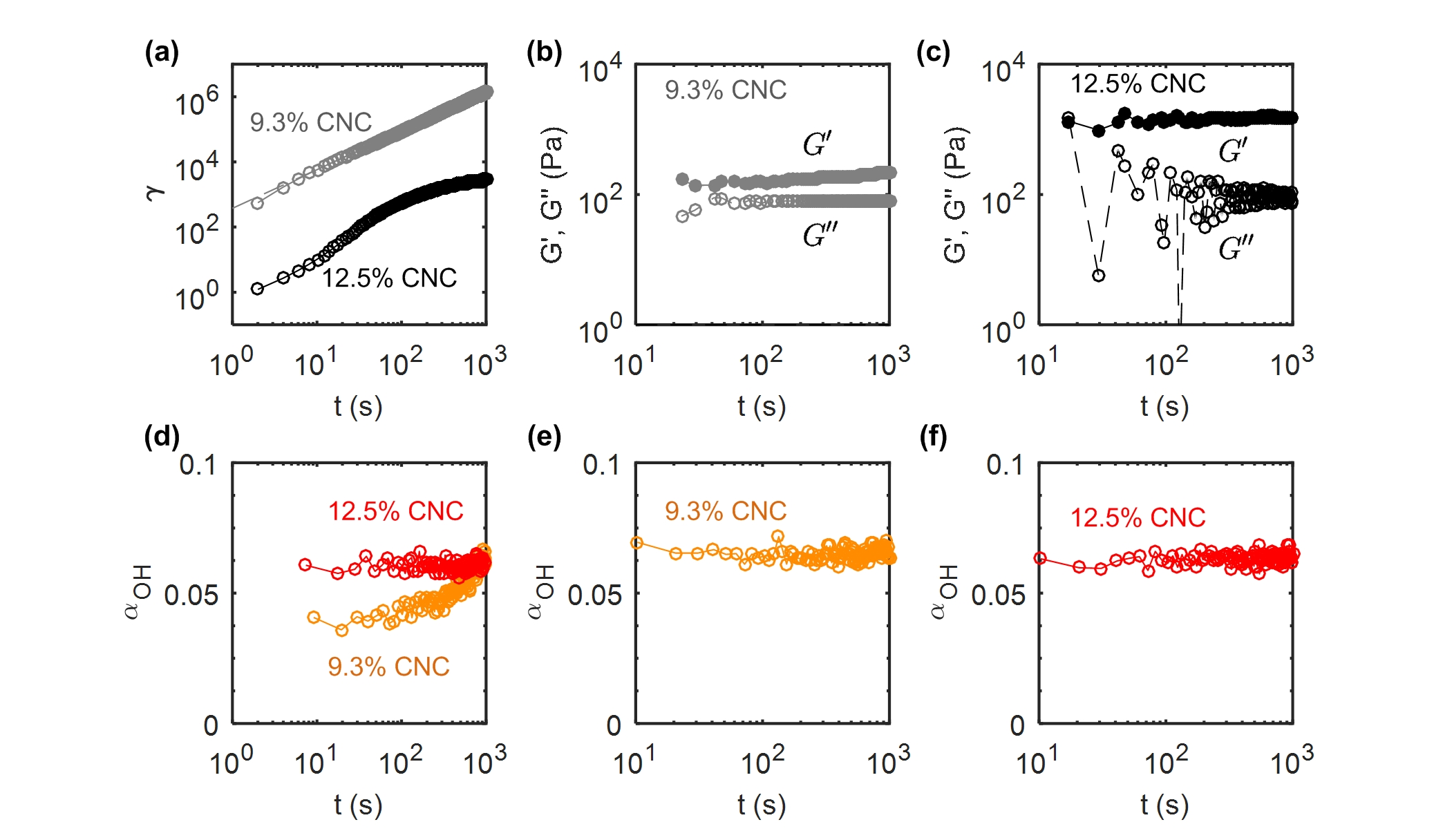}
    \caption{Creep and recovery of the composite (CG) gels: (a) Shear strain $\gamma$ versus time during creep flow at $\sigma=100$~Pa ($\sigma/\sigma_c = 1.71$) for a 9.3\% wt.~CNC gel and $\sigma=120$~Pa ($\sigma/\sigma_c = 0.86$) for a 12.5\%wt.~CNC gel. Refer to Fig.~\ref{UVCG_CG_yield_creep} in Appendix~\ref{appendix_yielding} for the determination of the yield stress. The dashed line corresponds to the best power-law fit of the data for the 9.3\% wt.~gel: $\gamma(t) \propto t^{1.19}$. (b) and (c) Elastic and viscous modulus versus time, measured by small amplitude oscillatory shear of amplitude $\sigma_0=0.5$~Pa and frequency $f=1$Hz for the 9.3\% wt.~and the 12.5\% wt.~CNC gel respectively. (d)--(f) Evolution of the IR peak ratio parameter $\alpha_{OH}$ versus time during (d) the creep flow and (e)--(f) the structural recovery of the 9.3\% wt.~and the 12.5\%~wt.~CNC gels showing an irreversible increase in the free OH peak.}
    \label{CG_creep}
    \end{center}
\end{figure*}

\subsection{Creep flows of composite gels}

While shear-rate controlled rheometry provides valuable insight into the irreversible shear melting of the gels, flow during processing of CNC composites often occurs under constant external stress, and therefore over different timescales. For instance, shear-induced rearrangements of CNCs during pressure-driven extrusion through a nozzle have been shown to take place over time scales up to 10$^3$~s \cite{Hausmann2018}. In a pressure-controlled extruder, the time over which a gel experiences a high shear stress is dependent on the geometry of the nozzle. Flow occurs when the stress at the wall of the nozzle reaches the yield stress \cite{Lewis2006a}. Therefore, a creep test performed with a rheometer, at a stress close to the yield stress, mimics the dynamic response of the gel during flow through an extruder. 

We first consider the case of the UV-curable gel at two different concentrations: 12\% wt.~and 14\% wt.~CNCs. The protocol is the following: the gel is subject to a 60~s pre-shear at $\dot \gamma=500~$s$^{-1}$ to erase the shear history from loading the gels into the rheometer. The gel is allowed to recover for 10$^3$~s, before being subjected to a stress sweep, which allows us to determine the yield stress of the gel (see supplementary Fig.~\ref{UVCG_CG_yield_creep}). The stress value, $\sigma$ used to perform the creep test is then chosen to be close to the yield stress, $\sigma/\sigma_c = 0.99$. The creep tests on the 12\% wt.~and 14\% wt.~gel were performed at $\sigma=65$~Pa and $\sigma=105$~Pa respectively, and the strain response is reported as a function of time Fig.~\ref{UV_CG_creep}a. 
The initial creep is slower for the 14\% wt.~gel compared to the 12\% wt. due to the higher elastic modulus of the material, but both gels accumulate a large total strain of about $\gamma_T \sim 10^6$ over 1000~s. The strain for the 12\%~wt.~CNC gel grows as a power-law of time: $\gamma(t) \sim t^{0.96}$, whereas the 14\%~wt.~gel shows a decelerating strain response. This result points toward a fluid-like behavior for the 12\%~wt.~gel and a more viscoelastic solid-like response for the 14\% wt.~gel.\cite{Chan2014} Indeed, this behavior is confirmed by the gels' response following the creep test (Fig.~\ref{UV_CG_creep}b and \ref{UV_CG_creep}c). Upon cessation of flow, the 12\%~wt.~gel is fully fluidized ($G'' < G'$), and the gel rebuilds within about 100~s. Note that both $G'$ and $G''$ experience an increase during the recovery. By contrast, the 14\%~wt.~gel shows a solid-like behaviour ($G'>G''$) immediately upon cessation of flow, which strongly suggests that the percolated network formed by the CNCs linked by hydrogen bonds still spans the whole sample, although the stress applied to the sample was larger than the yield stress. We attribute this difference to the denser microstructure of the gel.
Furthermore, the viscoelastic recovery for the 14\% wt.~gel only affects $G'$, while $G''$ is relatively constant, which suggests that the rebuilding of the percolated microstructure proceeds without an increase in the viscous dissipation. The corresponding changes in the hydrogen bond density between the CNCs are monitored with FTIR by following the evolution of $\alpha_{OH}$ during the creep and recovery tests. The results are consistent with the previous observations and confirm our rheological findings: $\alpha_{OH}$ increases monotonically during the creep test, i.e. hydrogen bonds between CNC particles are progressively destroyed under shear (Fig.~\ref{UV_CG_creep}d). Finally, $\alpha_{OH}$ remains constant during the recovery period, at a value equal to that reached at the end of the creep test, which means that the disruptions in the hydrogen bonds are not reversible during the rest period (Fig.~\ref{UV_CG_creep}e and \ref{UV_CG_creep}f). The elastic recovery in both 12\% wt.~and 14\% wt.~gels are therefore related to weak van der Waals attractive interactions between the CNCs. 

To examine the effect of charge-screening by the crosslinkers on the creep response of the gels, we repeated the same creep test on CNC composite gels without crosslinkers, at two different concentrations of 9.3\% wt.~and 12.5\% wt. In the absence of charge-screening, CNCs are stabilized by electrostatic repulsion, in addition to hydrogen bonds \cite{Xu2018}. Notably, the composite gels are observed to be translucent in comparison with the opaque white color of the UV-curable gels (Fig.~\ref{types_of_gels}). This is indicative of long-range order enabled by electrostatic interactions.
Creep experiments coupled to FTIR measurements were performed on the 9.3\% wt.~and the 12.5\% wt.~composite gels at $\sigma=100$~Pa and $\sigma=120$~Pa respectively. These stress values are again chosen to be slightly larger than the yield stress of the samples (see supplementary Fig.~\ref{UVCG_CG_yield_creep} for the determination of the yield stress). The creep response $\gamma(t,\sigma)$ of both gels are reported in Fig.~\ref{CG_creep}a on a logarithmic scale. Similar to the UV-curable gels, the strain of the 9.3\%~wt.~composite gel shows a power-law increase as a function of time: $\gamma(t)\sim t^{1.19}$, whereas the 12.5\% wt.~CNC gel shows a decelerating flow.  
 The key difference between the composite gel, which does not contain additives and the UV-curable gel (shown in Fig.~\ref{UV_CG_creep}) is visible in the response following the creep test (Fig.~\ref{CG_creep}b and \ref{CG_creep}c).
The absence of photoinitiator dramatically changes the recovery after flow. For both concentrations, $G'$ only weakly increases with time, while $G''$ remains constant. We attribute such a slow recovery of the composite gels to the presence of unscreened electrostatic interactions between the CNCs. 
Indeed, the electrostatic repulsion masks the attractive van der Waals interactions and thus prevents the aggregation of CNCs. Consequently, CNCs in these composite gels display cooperative and long-range slow motion, which accounts for the slow recovery. These features are the hallmark of glassy dynamics.\cite{Lu2013}
Therefore, our observations find a natural framework in the context of Soft Glassy Rheology, for which analogous effects have been reported, e.g; in laponite gels, where charge screening results in a transition from slow cooperative rearrangements of the particles, to rapid recovery \cite{Tanaka2005}. Finally for the 12.5\% wt.~composite gels, we note that $\alpha_{OH}$ remains constant during the creep test, due to the fact that that the stress selected for the creep test is not sufficiently higher than the yield stress (Fig.~\ref{CG_creep}d). As a result the total accumulated strain is insufficient to generate a significant change in $\alpha_{OH}$. However, the evolution of the strain still suggests that the sample is elastically deformed and flows largely as a plug (Fig.~\ref{CG_creep}a). Such a rheological response is favorable to retain the mechanical integrity of the gel during extrusion processing such as in direct-write printing. 

\section{Conclusions}
We have explored a set of formulations of CNC-polymer composite gels for the synthesis of UV-curable nanocomposites. This formulation involves the introduction of crosslinking additives, which confer tunable rheological properties upon the CNC gel, while enabling UV curing of highly filled CNC-polymer composites.  Dispersion of CNCs in DMF by sonication results in the formation of a metastable, time-evolving gel via hydrogen bonding between the CNCs. Using FTIR coupled with rheometry, we found that these hydrogen bonds are irreversibly broken by external shear, and that the density of hydrogen bonds increases monotonically with increasing cumulative strain. External shear thus plays the role of ``shear-melting" in CNC-polymer gels, and for sufficiently large strains, brings the gel into a fully fluidized state, which shows negligible recovery upon flow cessation. The presence of crosslinking additives screens the electrostatic repulsive interactions between the CNCs, which then interact mainly through van der Waals attractive interactions, enabling the UV-curable gels to recover their viscoelastic properties upon flow cessation. The elastic and viscous moduli of the UV-curable gels containing crosslinking additives display a power-law like recovery, which depends on the shear history and on the CNC concentration. Revealing universality of this behavior with respect to shear history and CNC concentration, we identified master curves that represent the recovery dynamics of the gels, for a broad range experimental parameters. Furthermore, the CNC composite gels display rheological phenomena similar to those observed in model colloidal gels, and furthermore, these behaviors can be tuned with additives such as crosslinking agents. This offers a unique opportunity to control the properties of CNC-polymer composite gels through applied shear, and for the application of these gels as precursors for solid nanocomposites. Therefore, this work should have a broad impact on the production of CNC-based materials, including by 3D printing.\\

\begin{acknowledgements}
The authors thank Nate Crawford (Thermo Fisher) for access to the Rheonaut instrument and assistance with the measurements; Crystal Owens for performing the DLS measurement; and S\'ebastien Manneville for insightful discussions. Financial support was provided by the Procter \& Gamble Company.  
\end{acknowledgements}


%

\section*{Supplemental Materials}

\section{CNC Size Distribution}
\label{appendix_dls}
\begin{figure}[h!]
    \centering
    \includegraphics[width=0.95\columnwidth]{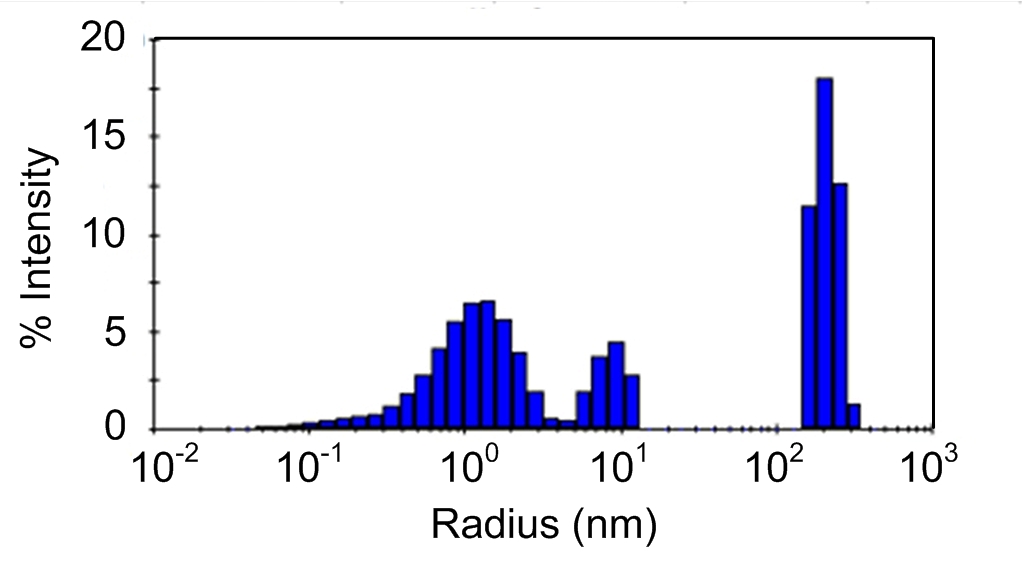}
    \caption{Size distribution of CNCs in an 8.5\% wt.~composite gel after probe sonication measured using DLS.}
    \label{cnc_size_dls}
\end{figure}

Dynamic light scattering (DLS) measurements were performed on 8.5\% wt.~CNC composite gels after the CNCs were dispersed by sonication. The weight fraction selected to ensure that the gels flow easily for insertion into the cuvettes for DLS measurements. The measurements were performed with a DynaPro Nanostar instrument. The aggregates in the freeze-dried CNC powder are dispersed into nanoparticles, with a majority of the particles having a diameter of about 2.5~nm. This corresponds to the expected diameter of individual CNCs. Figure \ref{cnc_size_dls} shows the DLS intensity distribution of the CNCs measured after probe sonication. The numerical description of the DLS peaks, including the mass distributions are provided in Table \ref{dls_table}.

\begin{table}[h!]
    \centering
    \begin{tabular}{ | m{1.4cm} | m{2cm}| m{1cm} | m{1.6cm} | m{1.4cm} |} 
    \hline 
    Item & Radius (nm) & PD \% & $M_{w}$ (kDa) & Mass \%\\ 
    \hline \hline 
    Peak 1 & 1.271 & 63.8 & 6 & 99.9  \\ 
    \hline
    Peak 2 & 8.933 & 22.8 & 565 & 0.1  \\ 
    \hline
    Peak 3 & 216.6 & 19.7 & 982087 & 0  \\ 
    \hline
    \end{tabular}
    \caption{Size distribution of CNCs after probe sonication measured using DLS.}
    \label{dls_table}
\end{table}

\section{Effect of crosslinkers}
\label{appendix_crosslinkers}
\begin{figure}
    \centering
    \includegraphics[width=0.95\columnwidth]{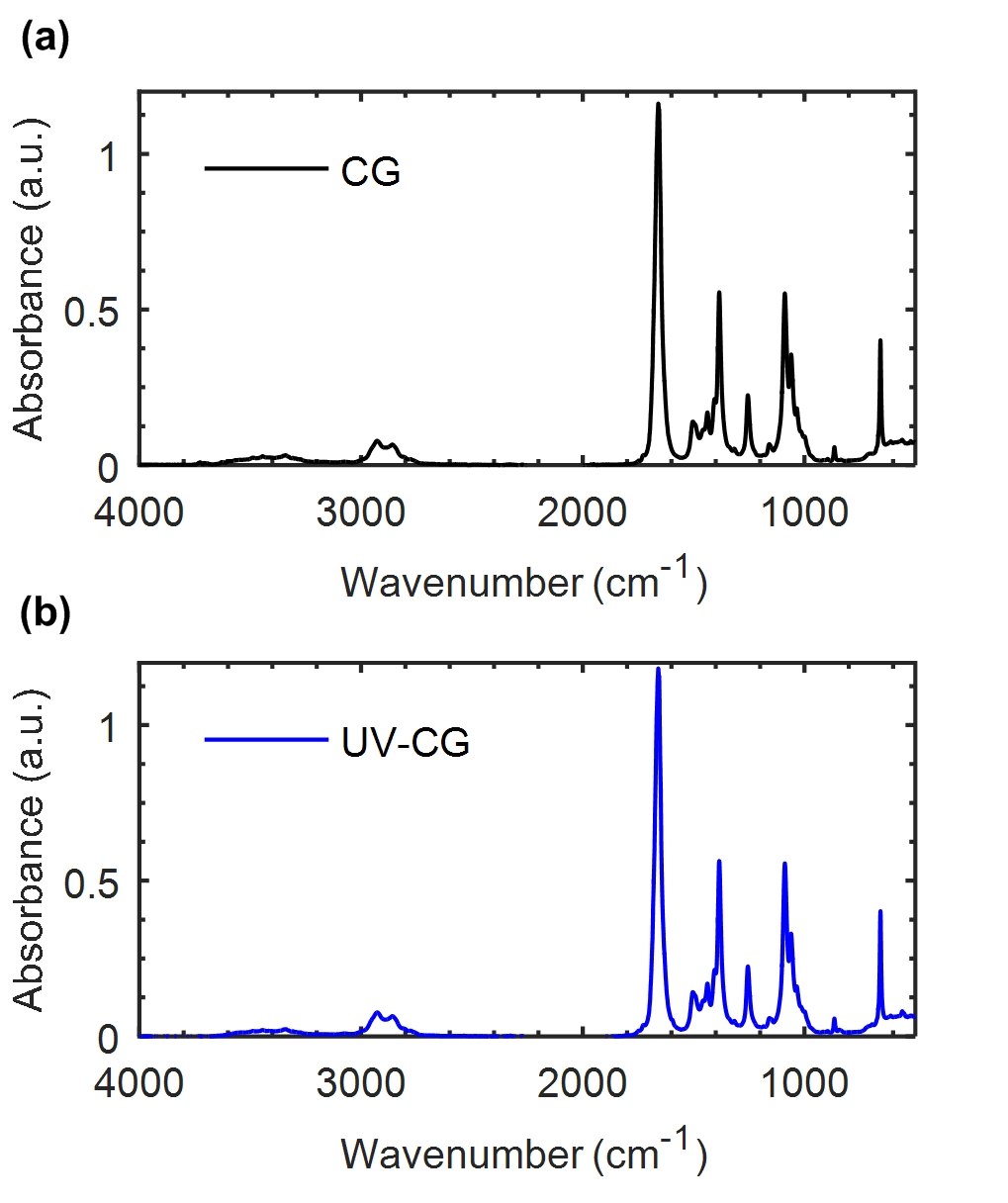}
    \caption{FTIR spectra of two types of CNC gels: (a) a composite gel (CG) with 12.5\% wt.~CNCs, and (b) a UV curable gel (UV-CG) with 12\% wt.~CNCs. The spectra for each sample are indistinguishable.}
    \label{uvcg_vs_cg_spectra}
\end{figure}

The composite CNC gels (CG) contain only the epoxide oligomer and solvent, while the UV-curable CNC gels (UV-CG) contain in addition a cationic photoinitiator and a polyamine thermal crosslinker. To examine whether the inclusion of the crosslinkers alters the physical interaction between the CNCs, or result in a chemical reaction, we compare the FTIR spectra of the two types of gels, prior to the application of any significant shear history. The composite CNC gel and the UV-curable CNC gel have 12.5\% wt.~and 12\% wt.~CNCs respectively (Fig.~\ref{uvcg_vs_cg_spectra}a and \ref{uvcg_vs_cg_spectra}b). The key peaks of interest in these two spectra are identical. These include the broad hydroxyl (OH) peak between 3200-3600~cm$^{-1}$, the CH peak around 2950~cm$^{-1}$, the COC bond due to the pyranose ring on cellulose at 1100~cm$^{-1}$ and the C=O peak from DMF located between 1600 and 1700~cm$^{-1}$. Therefore, we conclude that the crosslinkers do not have any discernable chemical effect on the gels.
The electrostatic repulsion between the CNCs is dependent on the Debye screening length of the medium \cite{Smith2016}. The presence of the thermal crosslinker and cationic photoinitiator modifies the nominal Debye screening length resulting in the inhibition of electrostatic repulsion between the CNCs. This allows the formation of clusters of CNC particles governed by van der Waals attraction in the UV-curable gels. 

\begin{figure*}
    \begin{center}
    \includegraphics[width=0.8\linewidth]{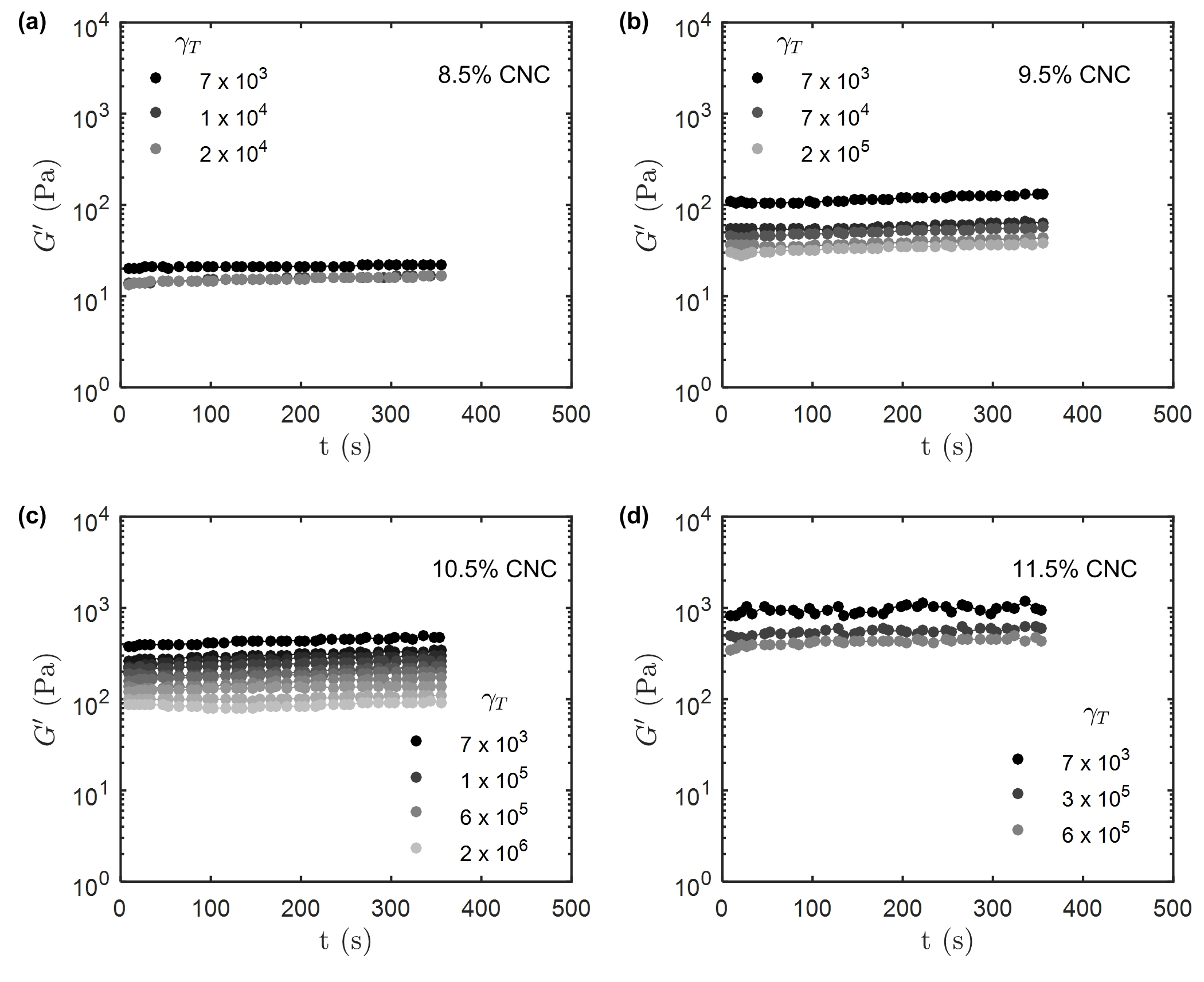}
    \caption{Absence of recovery of composite gels at different cumulative strains, after successive shear-melting steps indicated by increasing values of $\gamma_T$ for (a) 8.5\%, (b) 9.5\%, (c) 10.5\% and (d) 11.5\% wt.~CNC gels. The viscoelastic recovery was monitored by applying small amplitude oscillatory shear (stress amplitude $\sigma_0=0.5$~Pa and frequency $f=1$~Hz) following the same protocol as the one used for the UV curable gels (see main text Fig.~\ref{UV_CG_master_recovery}). To aid the quantitative comparison with Fig.~\ref{UV_CG_master_recovery} in the main text, the vertical scale of each graph is set to match that used for the UV curable gels.}
    \label{CG_PRT_protocol}
    \end{center}
\end{figure*}

\begin{figure}
    \centering
    \includegraphics[width=0.95\columnwidth]{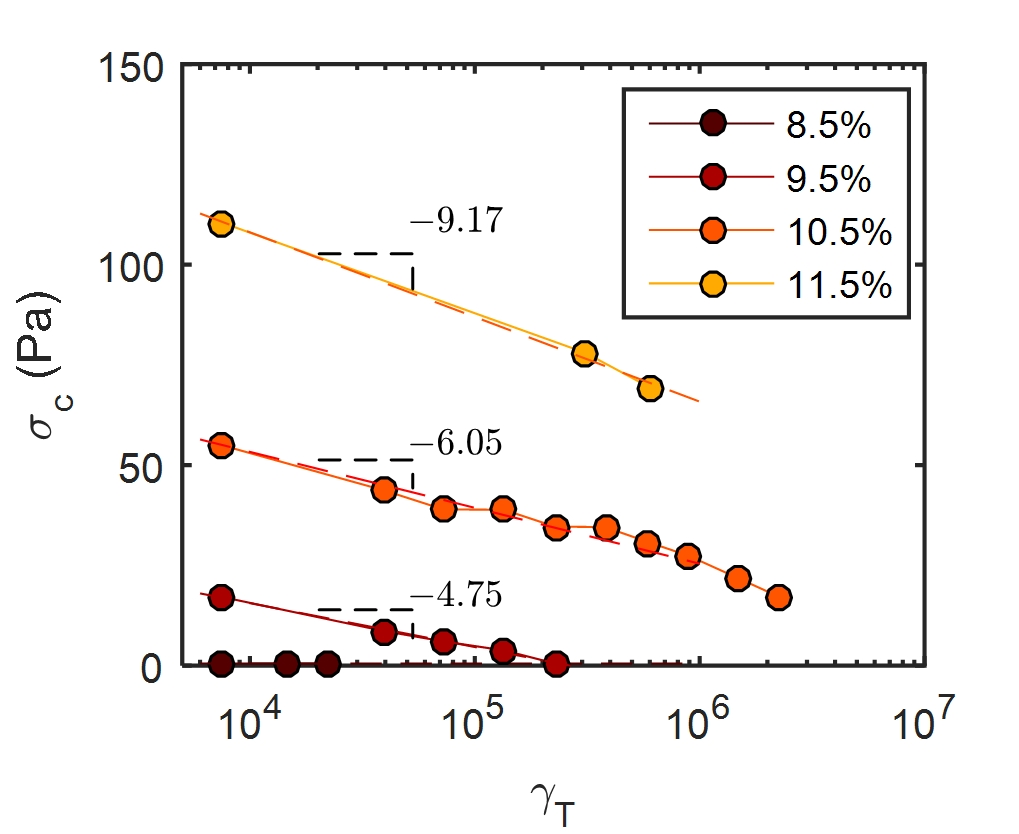}
    \caption{Yield stress $\sigma_c$ of composite gels for different CNC concentrations ranging between 8.5\% wt.~and 15\% wt.~vs the cumulative strain $\gamma_T$. Each yield stress was determined by performing a stress sweep at a frequency of $f=1$~Hz. The dashed lines correspond to the best fit of the data with a logarithmic function: $\sigma_c=\sigma_c^{(0)}+B\ln(\gamma_T)$, where $\sigma_c^{(0)}$ serve as free parameters. The fit parameter $B$ is reported on the graph.}
    \label{CG_yield_vs_gammaT}
\end{figure}

\section{Recovery of composite gels}
\label{appendix_CG_recovery}

The evolution of $G'$ versus time during the recovery of a composite gel, without crosslinkers, following the cessation of a shear-melting step is shown in Fig.~\ref{CG_PRT_protocol} for composite gels of different CNC concentrations. The yield stress determined after each recovery period is shown in Fig.~\ref{CG_yield_vs_gammaT}. For a detailed description of the rheology protocol, refer to Fig.~\ref{CSA166_PRT_consolidated}. In the absence of crosslinkers, the repulsive electrostatic interactions between the CNC particles prevent the formation of percolating CNC clusters. For 8.5\% wt.~composite gels, a small strain ($\gamma_T = 3\cdot 10^3$) is sufficient to fully and irreversibly fluidize the gel. The same finding applies to the 9.5\% wt.~CNC gel, which is irreversibly fluidized at $\gamma_T = 2\cdot10^5$. At higher concentrations of CNCs, i.e. (10.5\% wt.~and 11.5\% wt.), the elastic modulus progressively decreases with increasing strain, but the gel maintains a finite yield stress for  strains of at least $\gamma_T=6\cdot10^5$.

\begin{figure}[h!]
    \centering
    \includegraphics[width=0.95\columnwidth]{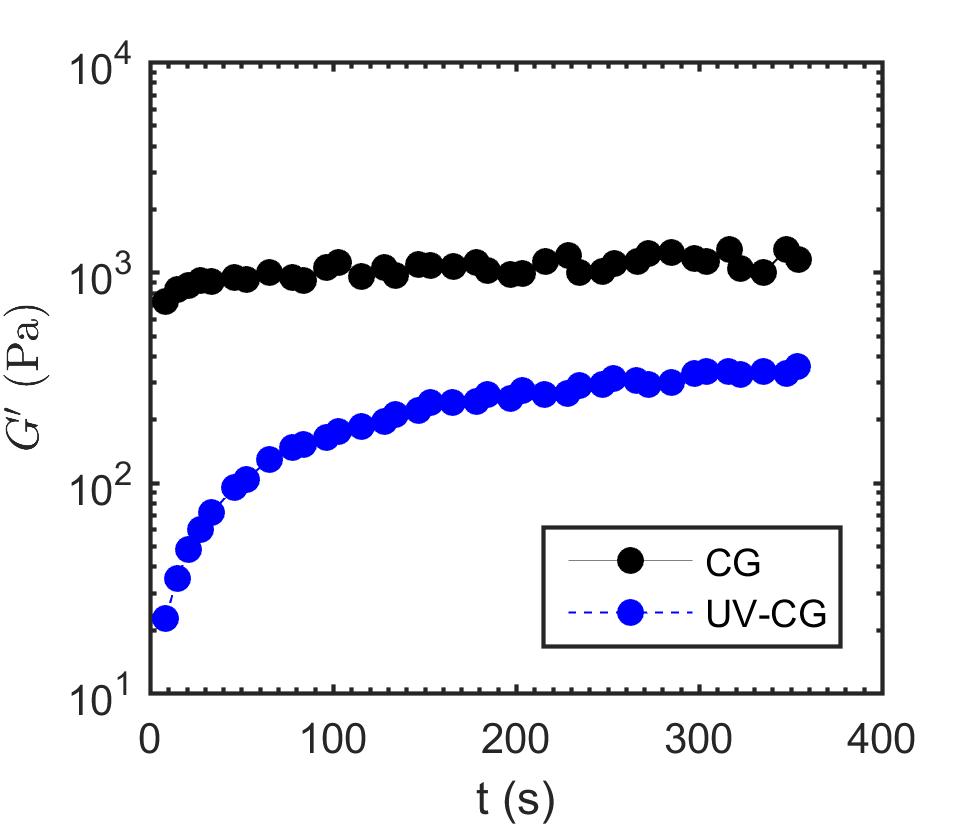}
    \caption{Comparison of the recovery of the elastic modulus of a 12.5\% wt.~composite gel and a 12\% wt.~UV-curable gel. The evolution of $G'$ for both types of gels were recorded after the application of a cumulative strain of $\gamma_T = 7\cdot 10^3$.}
    \label{P10_recovery_12pct_UVCG_vs_CG}
\end{figure}

The elastic properties of composite gels do not show the same recovery as seen in the UV-curable gels. During each recovery period, $G'$ shows a weak aging with time, with a relatively small change. Fig.~\ref{P10_recovery_12pct_UVCG_vs_CG} shows the recovery of the elastic modulus of a 12.\% wt.~composite gel (CG) with and a 12\% wt.~UV-curable gel (UV-CG) after the application of a shear strain of $\gamma_T = 7\cdot 10^3$. The 12\% wt.~UV-curable gel exhibits an initial elastic modulus $G'=22$ Pa, which recovers to about $G'=300$~Pa after 360~s. By contrast, for the same accumulated strain, a 12.5\% wt.~composite gel exhibits an initial elastic modulus $G'=818$~Pa, which increases to $G'=930$~Pa after 360~s. In the case of the composite gel, the presence of repulsive interactions in addition to attractive hydrogen bonds results in ``caging" of colloidal, which is responsible for the slow dynamics of recovery described above. 

\begin{figure}[h!]
    \centering
    \includegraphics[width=0.95\columnwidth]{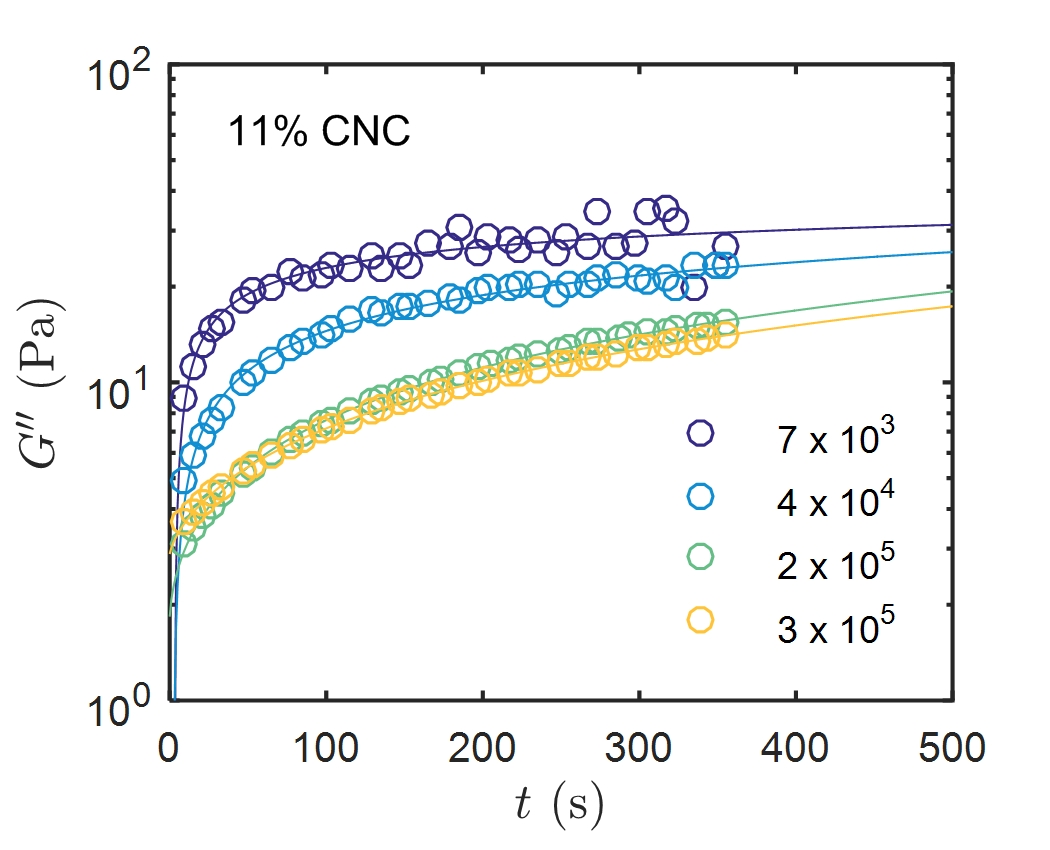}
    \caption{Recovery of the viscous modulus $G''$ versus time after shear-melting steps for a UV-curable gel with 12\% wt.~CNC. The cumulated strain experienced by the sample before each recovery is indicated on the graph. The fits corresponds to the best power-law fit of the data described by Eq.~\ref{power_law_aging} in the main text, in which $G'$ is replaced by $G''$}
    \label{CSA166_Gpp_recovery}
\end{figure}

\begin{figure}[h!]
    \centering
    \includegraphics[width=0.95\columnwidth]{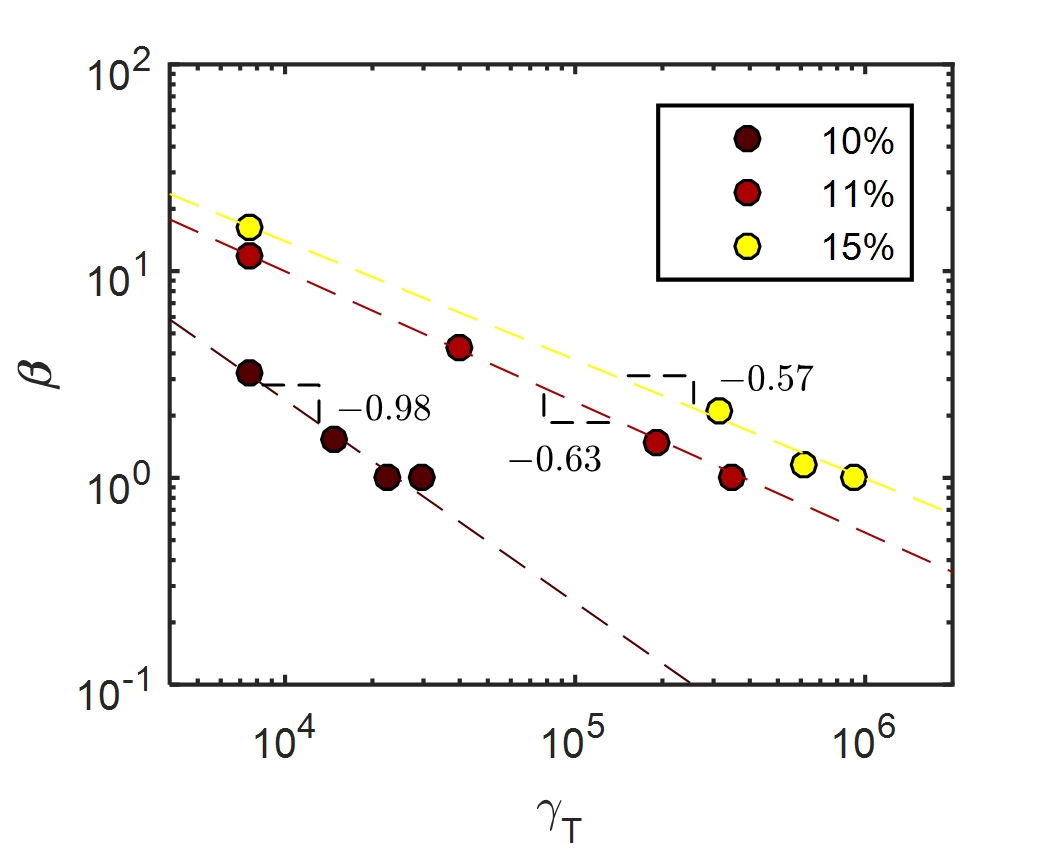}
    \caption{Dimensionless rescaling factor $\beta$ vs the cumulative strain $\gamma_T$ for UV-curable gels of three different concentrations: 10\% wt., 11\% wt. and 15\%~wt. (The corresponding rescaled curves are reported in Fig.~\ref{UV_CG_master_recovery}(b)--(d) in the main text.) The dashed ines correspond to the best power-law fit of the data: $\beta=k\gamma_T^{\zeta}$. The exponent $\zeta$ is reported on the graph.}
    \label{Beta_vs_gammaT}
\end{figure}

\section{Recovery of UV-curable gels}
\label{appendix_UVCG_details}
The recovery of UV-curable gels is described in Fig.~\ref{UV_CG_master_recovery} in the main text. Fig.~\ref{CSA166_Gpp_recovery} shows the evolution of the viscous modulus $G''$ with time for the recovery of a 11\% wt. UV curable gel as a complement to the evolution of $G'$ shown in Fig.~\ref{UV_CG_master_recovery}a. A power-law function identical to Eq.~\ref{power_law_aging} describes the evolution of $G''$ during the recovery. 

The  factor $\beta$ used to rescale the time axis of $G'(t)$ and $G''(t)$ for the 10\% wt., 11\% wt.~and 15\% wt.~UV-curable gels that are presented in Fig.~\ref{UV_CG_master_recovery}b--\ref{UV_CG_master_recovery}d  are reported in Fig.~\ref{Beta_vs_gammaT} as a function of the cumulative strain $\gamma_T$. For each UV-curable gel, the rescaling factor $\beta$ follows a power-law with respect to $\gamma_T$. The power-law exponent decreases for increasing CNC concentrations: from about $\zeta = -0.97$ for the 10\% wt.~gel to $\zeta = -0.57$ for the 15\% wt.~gel.

\begin{figure*}[ht!]
    \centering
    \includegraphics[width=0.8\linewidth]{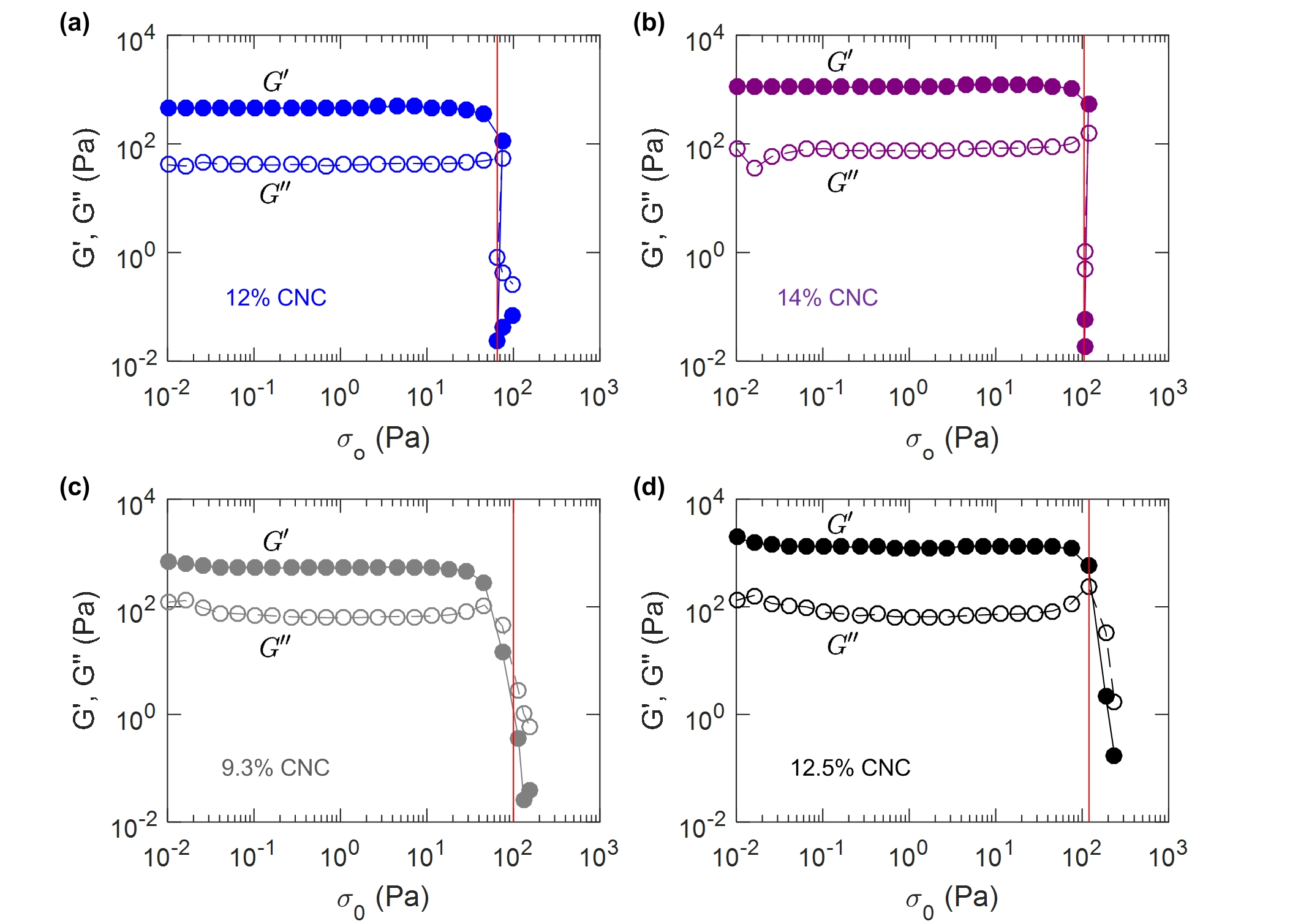}
    \caption{Stress sweeps on polymer-CNC gels showing $G'$ (filled symbols) and $G''$ (open symbols) versus the stress amplitude, $\sigma_0$. (a) 12\% wt.~and (b) 14\% wt.~UV-curable CNC gels.(c) 9.3\% wt.~and (d) 12.5\% wt.~composite CNC gel. Stress sweeps are performed at $f=1$~Hz following a 60~s pre-shear at $\dot \gamma=500$~s$^{-1}$ and a recovery period of 1000~s. The vertical red lines indicate the stresses used for the creep experiments described in Fig.~\ref{UV_CG_creep}a and Fig.~\ref{CG_creep}a.}
    \label{UVCG_CG_yield_creep}
\end{figure*}

\section{Comparison between the yielding behavior of composite gels and UV-curable gels}
\label{appendix_yielding}
The applied shear stress magnitude for the creep experiments described in Fig.~\ref{UV_CG_creep} and Fig.~\ref{CG_creep} were chosen in the vicinity of the yield stress, $\sigma_c$ of the sample at the present age. The latter value was determined by performing stress sweeps at $f=1$~Hz. The stress sweeps are reported in Fig.~\ref{UVCG_CG_yield_creep}a and \ref{UVCG_CG_yield_creep}b for the 12\% wt.~and 14\%~wt.~UV-curable gels, and in Fig.~\ref{UVCG_CG_yield_creep}c and \ref{UVCG_CG_yield_creep}d for the 9.3\% wt. and 12.5\% wt. composite gels. The vertical red lines indicate the stress at which the subsequent creep experiments were performed. During the stress sweep, the UV-curable gels show an abrupt yielding transition as both $G'$ and $G''$ drop by about two orders of magnitude within around 20~Pa of stress amplitude. By comparison, the yielding of the composite gels in the absence of charge screening crosslinkers, is observed to be more gradual. Around the yield stress, the elastic and viscous modulus show a smooth decay over about a decade of over a range of around 100~Pa in stress amplitude. 
he abrupt yielding of the UV-curable gels is reminiscent of a brittle-like failure scenario, whereas the progressive failure of the composite gels is more ductile-like. The study of large deformations of biopolymer gels with different intermolecular interactions shows that after yielding, the elastic modulus ''weak'' gels displays a significantly higher sensitivity to applied deformation, in comparison with ''strong'' gels \cite{Picout2003}. In the case of the composite gels the absence of charge screening creates additional repulsive interactions provide an additional entrapment effect for the CNCs, effectively resulting in a stronger gel.

\end{document}